\documentclass[twocolumn]{aastex702}

\usepackage{amsmath}
\usepackage{tabularx}
\usepackage{multirow}
\submitjournal{PASP}

\begin{document}

\title{Estimating M Dwarf Ages using their Transiting Brown Dwarf Companions}

\author[0009-0002-3373-9891]{Aryan Gupta}
\affiliation{Department of Astrophysical Sciences, Princeton University, Princeton, NJ 08540, USA}
\email{ary2027gupta@gmail.com}

\author[0000-0002-8965-3969]{Steven Giacalone}
\altaffiliation{NSF Astronomy and Astrophysics Postdoctoral Fellow}
\affiliation{Department of Astronomy, California Institute of Technology, Pasadena, CA 91125, USA}
\email{giacalone@astro.caltech.edu}

\author[0000-0002-4265-047X]{Joshua N.\ Winn}
\affiliation{Department of Astrophysical Sciences, Princeton University, Princeton, NJ 08540, USA}
\email{jnwinn@princeton.edu}

\author[0000-0003-0742-1660]{Gregory J. Gilbert}
\affiliation{Department of Astronomy, California Institute of Technology, Pasadena, CA 91125, USA}
\email{ggilbert@caltech.edu}

\author[0000-0003-4408-0463]{Zafar Rustamkulov}
\affiliation{IPAC, California Institute of Technology, MC 100-22, 1200 E. California Boulevard, Pasadena, CA 91125, USA}
\email{zafar@ipac.caltech.edu}

\author[0000-0002-9305-5101]{Luke B. Handley}
\affiliation{Department of Astronomy, California Institute of Technology, Pasadena, CA 91125, USA}
\email{lhandley@caltech.edu}

\author[0000-0001-8638-0320]{Andrew W. Howard}
\affiliation{Department of Astronomy, California Institute of Technology, Pasadena, CA 91125, USA}
\email{ahoward@caltech.edu}

\author[0000-0002-0672-9658]{Jonathan Gomez Barrientos}
\affiliation{Division of Geological and Planetary Sciences, California Institute of Technology, Pasadena, CA 91125, USA}
\email{jgomezba@caltech.edu}

\author[0000-0002-5375-4725]{Heather A. Knutson}
\affiliation{Division of Geological and Planetary Sciences, California Institute of Technology, Pasadena, CA 91125, USA}
\email{hknutso2@caltech.edu}

\author[0000-0001-9518-9691]{Morgan Saidel}
\affiliation{Carnegie Science Observatories, 813 Santa Barbara Street, Pasadena, CA 91101, USA}
\email{msaidel@carnegiescience.edu}

\begin{abstract}

Determining ages for M dwarfs remains challenging because of their slow evolution on the main sequence. We investigate using transiting brown dwarfs to constrain system age through their measured masses and thermal emission. We report Palomar/WIRC secondary-eclipse observations of two close-in brown dwarfs, LP~261-75C and AD~3116B, measuring eclipse depths of $6.1 \pm 0.3$~parts per thousand (ppt) in $K_s$ and $9.1 \pm 0.6$~ppt in $J$, respectively. Combining these observations with space-based transit photometry, radial velocities, and host star spectra, we infer the internal luminosities of these brown dwarfs and, assuming coevality, host star ages using five sets of atmospheric and evolutionary models. For LP~261-75C, the model-dependent median ages span 1.3--2.9~Gyr, substantially older than the previously proposed age of 0.1--0.2~Gyr based on presumed membership in the AB~Doradus moving group. This result, together with the host star's high metallicity and the companion's measured radius, disfavors membership in AB~Dor. The revised system age implies a mass of $\approx 62$--$68\,M_{\mathrm{Jup}}$ for the wide-orbit companion LP~261-75B, compared with an earlier estimate of $\approx 21\,M_{\mathrm{Jup}}$. AD~3116B, a member of the 0.6--0.8~Gyr Praesepe cluster, provides an opportunity to evaluate different brown dwarf evolutionary models as its mass, age, and luminosity can be independently constrained. Its inferred age depends strongly on the cloud treatment, spanning 0.4--0.6~Gyr for cloudy models and 1.0--1.3~Gyr for cloudless models. Among the models considered, the Sonora Diamondback model, which includes cloud condensation and cloud clearing across the L/T boundary, provides a plausible simultaneous match to the observed eclipse depth and radius. These results demonstrate a new methodology to constrain M dwarf ages using emission observations of their brown dwarf companions, while highlighting how inferred temperatures and ages are sensitive to evolutionary model differences.

\end{abstract}



\section{Introduction}

Brown dwarfs play an important role across several subdisciplines of astrophysics, including galactic archaeology \citep[e.g.,][]{burgasser2003,burgasser2004,zhang2017,aganze2022}, stellar evolution \citep[e.g.,][]{chabrier2000,burrows2001,dieterich2014,kirkpatrick2021}, and exoplanet atmospheres \citep[e.g.,][]{marley2015,faherty2016,miles2023,beiler2023}. The usefulness of brown dwarfs for these applications depends on accurate models of their atmospheric chemistry and their thermal evolution over billions of years. Many such models have been produced in the last few decades, each of which has added new dimensions of complexity needed to reproduce observations, such as improved equations of state, new prescriptions for vertical atmospheric mixing, and cloud physics \citep[e.g.,][]{burrows1997,baraffe2003,saumon2008,phillips2020,marley2021}. These models describe the evolutionary state of a brown dwarf using three parameters: mass, luminosity, and age. When two of these quantities are known, the third can be estimated.

In this paper, we explore a new potential application of brown dwarf evolutionary models: estimating the ages of systems consisting of M dwarfs and bound brown dwarf companions (dM+BD). M dwarfs, which constitute roughly $70\%$ of stars in the Galaxy, are notoriously challenging to age-date due to their very gradual evolutions \citep{henry2006,soderblom2010,reyle2021}. Establishing new ways to determine the ages of these low-mass stars would open new pathways for understanding their interiors, their activity states, and the properties of planets that orbit them \citep[e.g.,][]{west2008,newton2016mdwarf,shields2016}. Transiting brown dwarfs offer a unique opportunity in this respect; their masses can be measured precisely using radial velocity observations and their luminosities can be determined from the emergent flux measured from secondary eclipse observations. With both mass and luminosity in hand, one can estimate the ages of the brown dwarfs and their host stars, assuming they are coeval.

Currently, M dwarf age dating is limited to only a few techniques. The largest fraction of M dwarfs with precisely constrained ages come from kinematic associations. For instance, most M dwarfs with constrained ages younger than $\approx$ 1~Gyr belong to stellar clusters whose ages have been determined by fitting the observed HR diagram with stellar-evolutionary models \citep[e.g.,][]{bell2015,douglas2019praesepe,cantatgaudin2020,engle2023}. At the other extreme, some of the oldest known M dwarfs ($\gtrsim 8$~Gyr) have been age-dated due to membership in the galactic thick disk or halo \citep[e.g.,][]{bochanski2007,lepine2007,newton2016mdwarf,engle2023}. For M dwarfs with intermediate ages, the age is challenging to determine observationally. Age determinations typically rely on the presence of a bound companion whose age can be measured via other techniques. For instance, M dwarfs with white dwarf companions can be age dated using empirically calibrated white dwarf cooling models \citep[e.g.,][]{garces2011,kiman2021,chiti2024}. In a few cases, ages have been measured for M dwarfs with main-sequence companions by constraining the age of the primary star with gyrochronology or asteroseismology \citep[e.g.,][]{chontos2021,pass2022}. In recent years, these measurements have enabled the first studies of M dwarf spin down, providing a path towards estimating M dwarf ages based on their rotation periods \citep{newton2016mdwarf,pass2022,pass2024,engle2023,chiti2024,lu2024}. However, M dwarfs with intermediate, well-constrained ages still remain sparsely sampled. If dM+BD systems could be age-dated using brown dwarf evolutionary models, it would improve the prospects of accurately calibrating M dwarf spin down models. In this respect, benchmark brown dwarfs -- those in which their luminosities, masses, and ages have all been determined independently of the evolutionary models \citep[e.g.,][]{pinfield2006,dupuy2009,dupuy2017,bowler2018benchmark,brandt2019benchmark,rickman2020benchmark,brandt2021benchmark,bonavita2022benchmark,rickman2024benchmark,li2026benchmark} -- will be crucial for determining the efficacy of any dM+BD age-dating method.

In this paper, we estimate the ages of two M dwarfs with transiting brown dwarf companions, LP~261-75C and AD~3116B, the latter of which is now a benchmark system due to its independently measured age. The outline of this paper is as follows. In Section~\ref{sec:targets}, we describe the two systems targeted in this paper. In Section~\ref{sec:observational_data}, we describe the data used in our analysis. In Section~\ref{sec:analysis}, we detail our analysis procedure, including our joint fit of the photometry and radial velocities and our estimation of stellar ages for different brown dwarf evolutionary models. In Section~\ref{sec:results}, we present our age results, test different evolutionary models using our benchmark system AD~3116, and discuss our findings. In Section~\ref{sec:future}, we discuss the prospects of measuring M dwarf ages using brown dwarf companions in the future. In Section~\ref{sec:conclusions}, we present a summary of our findings and our conclusions.

\section{Targets}\label{sec:targets}

\subsection{LP~261-75} \label{sec:2pt1}

The LP~261-75 system consists of a mid-type M dwarf host star (LP~261-75A), a wide-separation brown dwarf companion (LP~261-75B; \citealt{reid2006lp26175}), and a more recently discovered brown dwarf that transits the M dwarf (LP~261-75C; \citealt{irwin2018lp26175}). LP~261-75B is directly imaged, has a projected separation of $450 \pm 120$~au, and was determined to be bound to the M dwarf primary based on common proper motions. This wide-separation companion has a spectral type of L6 ($T_{\rm eff} = 1500 \pm 150 \, {\rm K}$) and an estimated mass of $M_{\rm B} = 21^{+10}_{-5} \, M_{\rm Jup}$ (assuming an age of $100-200$~Myr; \citealt{reid2006lp26175}). LP~261-75C has an orbital period of $P_{\rm orb} = 1.88$~days around star A, a radius of $R_{\rm C} = 0.90 \pm 0.02 \, R_{\rm Jup}$, and a mass of $M_{\rm C} = 67.4 \pm 2.1 \, M_{\rm Jup}$ \citep{irwin2018lp26175}. Our adopted stellar parameters for
star A are shown in Table~\ref{tab1}.

The age of the LP~261-75 system is uncertain. \citet{reid2006lp26175} noted that the chromospheric activity levels of the M dwarf are consistent with similar M dwarfs in the $100-200$~Myr Pleiades cluster (also see \citealt{shkolnik2009young}). Using ground-based photometry, \citet{irwin2018lp26175} measured the rotation period of the M dwarf to be $P_{\rm rot} \approx 2.22$~days (later confirmed by \citealt{brady2025maroonx} using TESS photometry), which is typical for young M dwarfs. However, \citet{irwin2018lp26175} urged caution in interpreting the age based on chromospheric activity and rotation period alone, given that M dwarfs have been observed to remain active and rapidly rotating for longer than $1$~Gyr \citep[e.g.,][]{newton2016mdwarf} and because tidal interactions between the star and its close-in brown dwarf companion have likely inhibited stellar spin down. 

\citet{irwin2018lp26175} also used the kinematics of the system to estimate its age, finding that it is a close kinematic match to the $100-200$~Myr AB Doradus moving group. However, they acknowledged that the interpretation of this finding is complicated by the fact that stars are often misidentified as AB Dor members based on kinematics alone. For instance, many AB Dor ``stream'' stars do not share the same ages and metallicities as those originating from the main AB Dor ``nucleus'' \citep[e.g.,][]{barenfeld2013abdor}. Indeed, our near-infrared spectroscopic analysis of LP~261-75A yielded ${\rm [Fe/H]}=+0.52\pm0.09$ and $+0.59\pm0.08$ from early- and late-M-dwarf metallicity calibrations (see Section~\ref{sec:irtf}), indicating a high metallicity of ${\rm [Fe/H]}\gtrsim+0.5$, which is discrepant with the near-solar metallicities measured for  AB Doradus members \citep{barenfeld2013abdor, maccarthy2014abdor}.  Furthermore, a more up-to-date kinematic analysis using BANYAN $\Sigma$ \citep{gagne2018banyan} and data from Gaia DR3 \citep{gaia2023gaiadr3} found a $99.9 \%$ probability of LP~261-75A being a field star. Lastly, \citet{brady2025maroonx} noted that the radius of the brown dwarf is smaller than evolutionary models predict it should be for an age of $100-200$~Myr. Taken together, these measurements suggest that this system does not belong to AB Dor and is in fact much older than the members of AB Dor.

\subsection{AD~3116}\label{ad3116_subsection}

\begin{deluxetable}{lcc}[htbp]
\centering
\tabletypesize{\scriptsize}
\tablewidth{0pt}
\tablecaption{Adopted LP~261-75A Stellar Parameters\label{tab1}}
\tablehead{\colhead{Parameter} & \colhead{Value} & \colhead{Source} }
\startdata
\multicolumn{3}{c}{Identifiers}\\
\hline
LP & 261-75 & \\
TOI & 1779 & \\ 
TIC & 67646988 &   \\
Gaia DR3 & 796443724462555904 &  \\
2MASS & J09510459$+$3558098 & \\
\hline
\multicolumn{3}{c}{Astrometry \& Position}\\
\hline
$\alpha$  & 09:51:04.44 & Gaia DR3 \\
$\delta$  & +35:58:06.72 & Gaia DR3 \\ 
$\mu_\alpha$ (mas\,yr$^{-1}$)& $-100.90 \pm 0.02$	& Gaia DR3\\
$\mu_\delta$  (mas\,yr$^{-1}$) & $-171.57 \pm 0.02$  & Gaia DR3\\
$\pi$ (mas) & $29.42\pm0.02$ & Gaia DR3\\
Distance (pc) & $33.99 \pm 0.02$ & From parallax \\
\hline
\multicolumn{3}{c}{Photometry}\\
\hline
${\rm TESS}$ & $12.437 \pm 0.007$ & TICv8.2 \\
$G$ & $13.804 \pm 0.001$ & Gaia DR3\\
$G_{\rm BP}$ & $15.616\pm0.001$ & Gaia DR3 \\
$G_{\rm RP}$ & $12.527\pm0.001$ & Gaia DR3 \\
$B$ & $16.9\pm0.2$ & APASS DR9\\
$V$ & $15.44\pm0.04$ & APASS DR9\\
$u$ & $18.50 \pm 0.02$ & SDSS DR8 \\
$g$ & $16.200 \pm 0.003$ & SDSS DR8 \\
$r$ & $14.727 \pm 0.003$ & SDSS DR8 \\
$i$ & $13.543 \pm 0.002$ & SDSS DR8 \\
$z$ & $12.154 \pm 0.003$ & SDSS DR8 \\
$J$ & $10.58\pm0.02$ & 2MASS \\
$H$ & $9.96\pm0.02$ & 2MASS \\
$K_s$ & $9.69\pm0.02$ & 2MASS \\
$W1$ & $9.53\pm0.02$ & ALLWISE \\
$W2$ & $9.37\pm0.02$ & ALLWISE \\
$W3$ & $9.18\pm0.02$ & ALLWISE \\
\hline
\multicolumn{3}{c}{Stellar Physical Properties}\\
\hline
$M_\star$ ($M_\odot$)& $0.28\pm0.02$& TICv8.2 \\
$R_\star$ ($R_\odot$)& $0.306\pm0.009$ & TICv8.2 \\
$\rho_\star$ (${\rm g} \, {\rm cm}^{-3}$) & $13.9 \pm 0.3$ & TICv8.2 \\
$T_{\rm eff}$ (K) & $3138\pm157$ & TICv8.2 \\
$L_\star$ ($L_\odot$) & $0.008\pm0.002$ & TICv8.2 \\
${\rm [Fe/H]}$ (dex) & $\gtrsim + 0.5$ & This paper \\
\enddata
\tablecomments{References for the sources listed above are as follows: TICv8.2 \citep{dressing2018cooldwarfs, stassun2019tesscatalogue}, Gaia DR3 \citep{gaia2023gaiadr3}, APASS DR9 \citep{henden2016apass}, SDSS DR8 \citep{aihara2011sdssdr8}, 2MASS \citep{skrutskie20062mass}, ALLWISE \citep{cutri2014allwise}. Note that the ALLWISE $W4$ magnitude has no uncertainty estimation, so we have omitted it from this table.}
\end{deluxetable}

\begin{deluxetable}{lcc}[htbp]
\centering
\tabletypesize{\scriptsize}
\tablewidth{0pt}
\tablecaption{Adopted AD~3116A Stellar Parameters\label{tab2}}
\tablehead{\colhead{Parameter} & \colhead{Value} & \colhead{Source} }
\startdata
\multicolumn{3}{c}{Identifiers}\\
\hline
AD & 3116 & \\
TOI & 5090 & \\ 
TIC & 184892124 &   \\
EPIC & 211946007 & \\
Gaia DR3 & 661229701489213824 &  \\
2MASS & J08423943$+$1924520 & \\
\hline
\multicolumn{3}{c}{Astrometry \& Position}\\
\hline
$\alpha$  & 08:42:39.40  & Gaia DR3 \\
$\delta$  & +19:24:51.69 & Gaia DR3 \\ 
$\mu_\alpha$ (mas\,yr$^{-1}$)& $-36.6\pm0.1$	& Gaia DR3\\
$\mu_\delta$  (mas\,yr$^{-1}$) & $-13.13\pm0.07$  & Gaia DR3\\
$\pi$ (mas) & $5.35\pm0.09$ & Gaia DR3\\
Distance (pc) & $187\pm3$ & From parallax \\
\hline
\multicolumn{3}{c}{Photometry}\\
\hline
${\rm TESS}$ & $16.057 \pm 0.008$ & TICv8.2 \\
$G$ & $17.378 \pm 0.001$ & Gaia DR3 \\
$G_{\rm BP}$ & $19.114\pm0.0001$ & Gaia DR3 \\
$G_{\rm RP}$ & $16.142\pm0.0001$ & Gaia DR3 \\
$u$ & $22.3 \pm 0.2$ & SDSS DR8 \\
$g$ & $19.65 \pm 0.01$ & SDSS DR8 \\
$r$ & $18.212 \pm 0.007$ & SDSS DR8 \\
$i$ & $16.679 \pm 0.005$ & SDSS DR8 \\
$z$ & $15.853 \pm 0.006$ & SDSS DR8 \\
$J$ & $14.34\pm0.03$ & 2MASS \\
$H$ & $13.77\pm0.04$ & 2MASS \\
$K_s$ & $13.50\pm0.04$ & 2MASS \\
$W1$ & $13.33\pm0.03$ & ALLWISE \\
$W2$ & $13.11\pm0.03$ & ALLWISE \\
\hline
\multicolumn{3}{c}{Stellar Physical Properties}\\
\hline
$M_\star$ ($M_\odot$)& $0.27\pm0.02$& TICv8.2 \\
$R_\star$ ($R_\odot$)& $0.29\pm0.01$ & TICv8.2\\
$\rho_\star$ (${\rm g} \, {\rm cm}^{-3}$) & $15.1 \pm 0.6$ & TICv8.2 \\
$T_{\rm eff}$ (K) & $3213\pm157$ & TICv8.2 \\
$L_\star$ ($L_\odot$) & $0.008\pm0.002$ & TICv8.2 \\
Age (Myr) & $600-800$ & Various  \\
\enddata
\tablecomments{References for the sources listed above are as follows: TICv8.2 \citep{dressing2018cooldwarfs, stassun2019tesscatalogue}, Gaia DR3 \citep{gaia2023gaiadr3}, SDSS DR8 \citep{aihara2011sdssdr8}, 2MASS \citep{skrutskie20062mass}, ALLWISE \citep{cutri2014allwise}. Note that the ALLWISE $W3$ and $W4$ magnitudes have no uncertainty estimations, so we have omitted them from this table.}
\end{deluxetable}

The AD~3116 system consists of a mid-type M dwarf host star (AD~3116A) and a close-in transiting brown dwarf companion (AD~3116B; \citealt{gillen2017ad3116}). AD~3116B has an orbital period of $P_{\rm orb} = 1.98$~days, a radius of $R_{\rm B} = 0.95 \pm 0.07 \, R_{\rm Jup}$, and a mass of $M_{\rm B} = 54.6 \pm 6.8 \, M_{\rm Jup}$ \citep{carmichael2023bds}. Our adopted stellar parameters for the host star are shown in Table~\ref{tab2}.

According to \citet{kraus2007praesepe}, the AD~3116 system has a $99.1 \%$ probability of membership in the Praesepe open cluster based on an analysis of its proper motion and broad-band photometry. \citet{gillen2017ad3116} reported a stellar rotation period of $P_{\rm rot} = 2.25 \pm 0.02$~days based on light curves from the NASA K2 mission \citep{howell2014k2}, which supports the notion that this system is relatively young. However, like LP~261-75, we note that tidal interactions have likely inhibited spin down in this system, making rotation rate a poor tracer of system age. Characterizations of Praesepe vary, but most produce ages within the range of $600 - 800$~Myr \citep{salaris2004praesepe, gaspar2009praesepe, bell2014praesepe, brandt2015praesepe, gossage2018praesepe, douglas2019praesepe, salaris2019praesepe, morales2022praesepe, alfonso2023praesepe, pamos20233praesepe} and metallicities within the range of $[\mathrm{Fe/H}]\simeq+0.1$ -- $+0.3$ \citep{Pace2008,Boesgaard2013,DOrazi2020}. Because the age of the cluster is known independently from substellar cooling tracks and because the dynamical mass of the brown dwarf companion has been measured with high precision, AD~3116 is a rare benchmark system for testing brown dwarf evolutionary models \citep[e.g.,][]{cheetham2018benchmark, bowler2018benchmark, brandt2019benchmark, brandt2021benchmark, maire2020benchmark, rickman2020benchmark, rickman2024benchmark, bonavita2022benchmark, franson2022benchmark, franson2023benchmark, li2026benchmark}.

\section{Observations}\label{sec:observational_data}

\subsection{Palomar/WIRC Photometry}

We observed a single secondary eclipse of LP~261-75C and two secondary eclipses of AD~3116B with the Wide-field Infrared Camera (WIRC) on the 200-inch Hale Telescope at Palomar Observatory \citep{wilson2003wirc}. Observations of LP~261-75C were acquired on 2026-01-12 UT with the $K_s$ filter and observations of AD~3116B were acquired on 2026-03-21 UT and 2026-03-23 UT with the $J$ filter. All observations were taken under clear and dry conditions with the exception of those on 2026-03-21 UT, during which there was mild extinction from cirrus clouds. All observations were performed with a custom beam-shaping diffuser, which produces a top-hat-like point-spread function with a full width at half maximum of $3\arcsec$, designed to improve photometric precision for bright objects \citep{stefansson2017diffuser}. We reduced our observations using the publicly available \texttt{exowirc} pipeline \citep{vissapragada2020exowirc}, which performs dark subtraction, flat fielding, background subtraction, and dead/hot pixel correction using calibrations obtained at the beginning and end of the night. We generated our light curves using the built-in functionality of \texttt{exowirc}, which performs differential photometry between the target star and several nearby stars within the $8 \farcm 7$ field of view, selecting the optimal aperture size by minimizing the root mean square variability of the out-of-eclipse portion of the target star light curve. Systematics were detrended from each target light curve by fitting the product of a trapezoidal eclipse model and the weighted linear combination of comparison star light curves, airmass, point-spread-function width, and sky background. These fits were performed via MCMC using \texttt{PyMC3} with uniform priors on eclipse timing and depth \citep{salvatier2016probabilistic}. We used the resulting light curves in the global joint fit described in Section~\ref{sec:global_fits}.


\subsection{TESS Photometry}

Our analysis makes use of data from the NASA Transiting Exoplanet Survey Satellite (TESS; \citealt{ricker2015tess}). For both targets, we used 120-second-cadence Presearch Data Conditioning Simple Aperture Photometry (PDCSAP; \citealt{smith2012,stumpe2012,stumpe2014}) light curves generated by the NASA Science Processing Operations Center (SPOC) pipeline \citep{jenkins2016}. For LP~261-75, we used the data from TESS sectors 21 and 48. For AD~3116, we used the data from TESS sectors 44, 45, 46 and 72. These light curves were accessed from the Mikulski Archive for Space Telescopes (MAST) using the \texttt{lightkurve} Python package \citep{lightkurve2018}. For each target, we rejected all data points with NaN values, normalized each sector individually by dividing by the median flux, and masked outlier data (i.e., all data with normalized fluxes above 1.06 for LP~261-75 and outside the range of $0.5-1.5$ for AD~3116). We used these light curves in the global joint fit described in Section~\ref{sec:global_fits}.

\subsection{K2 Photometry}

Our analysis makes use of data from the NASA K2 mission \citep{howell2014k2}. AD~3116 was observed in K2 campaigns 5, 16, and 18, and the data are available with an 1800-second cadence. We used the K2SFF light curves, which were processed using the procedure described by \cite{vanderburg2014k2sff}. All K2 data were accessed from MAST using the \texttt{lightkurve} Python package \citep{lightkurve2018}. Because these light curves did not provide flux uncertainties, we estimated a constant uncertainty for each campaign from the scatter in the data. The light curves exhibited clear rotational modulation, so we first subtracted a Gaussian process fit with a Mat\'{e}rn kernel before computing the median absolute deviation (MAD) of the residuals. We then adopted $\sigma_{\mathrm{phot}}=1.4826\,\mathrm{MAD}$ as the photometric uncertainty. We normalized the data from each campaign individually and masked outlier data (flux values outside the range of $0.75-1.1$). We used these light curves in the global joint fit described in Section~\ref{sec:global_fits}.

\subsection{FLWO/TRES Radial Velocities}

Precise radial velocities of LP 261-75A were obtained with the Tillinghast Reflector Echelle Spectrograph (TRES; \citealt{szentgyorgyi2007tres}) on the 1.5-meter Tillinghast reflector at Fred Lawrence Whipple Observatory (FLWO), on Mt.\ Hopkins, Arizona, in order to measure the mass of the transiting brown dwarf. These data were first reported by \citet{irwin2018lp26175}. We used these radial velocities in the global joint fit described in Section~\ref{sec:global_fits}.

\subsection{Keck/HIRES Radial Velocities}

Precise radial velocities of AD~3116A were obtained with the HIRES spectrograph \citep{vogt1994hires} on the 10-meter Keck-I Telescope, on Maunakea, Hawaii, in order to measure the mass of the transiting brown dwarf. These data were first reported by \citet{gillen2017ad3116}. We used these radial velocities in the global joint fit described in Section~\ref{sec:global_fits}.

\subsection{Gemini/MAROON-X Radial Velocities}

LP~261-75C was observed on 2024-04-17 UT with the MAROON-X spectrograph \citep{seifahrt2018maroonx, seifahrt2022maroonx} on the 8.1-meter Gemini-North Telescope by \citet{brady2025maroonx}. In their paper, these precise, time-series radial velocities were used to measure the stellar obliquity of the system. We used these radial velocities in the global joint fit described in Section~\ref{sec:global_fits}.


\subsection{SPHEREx Spectroscopy}\label{sec:3pt7}

Both LP~261-75A and AD~3116A were observed by the NASA SPHEREx mission, which is collecting low-resolution ($R \approx 40 - 130$) spectra between $0.75-5.0 \, \mu{\rm m}$ of more than 100~million stars across the full sky \citep{crill2020spherex, bock2026spherex}. The data were reduced by the SPHEREx pipeline \citep{akeson2025spherex} and the spectra were extracted using the procedure outlined by \citet{rastumkulov2026spherex}. We used the SPHEREx spectra of the M dwarfs to estimate the intrinsic temperature of the transiting brown dwarf companions; see Section~\ref{sec:internal_temp}. 

\begin{figure*}[t!]
    \centering
    \includegraphics[width=\textwidth]{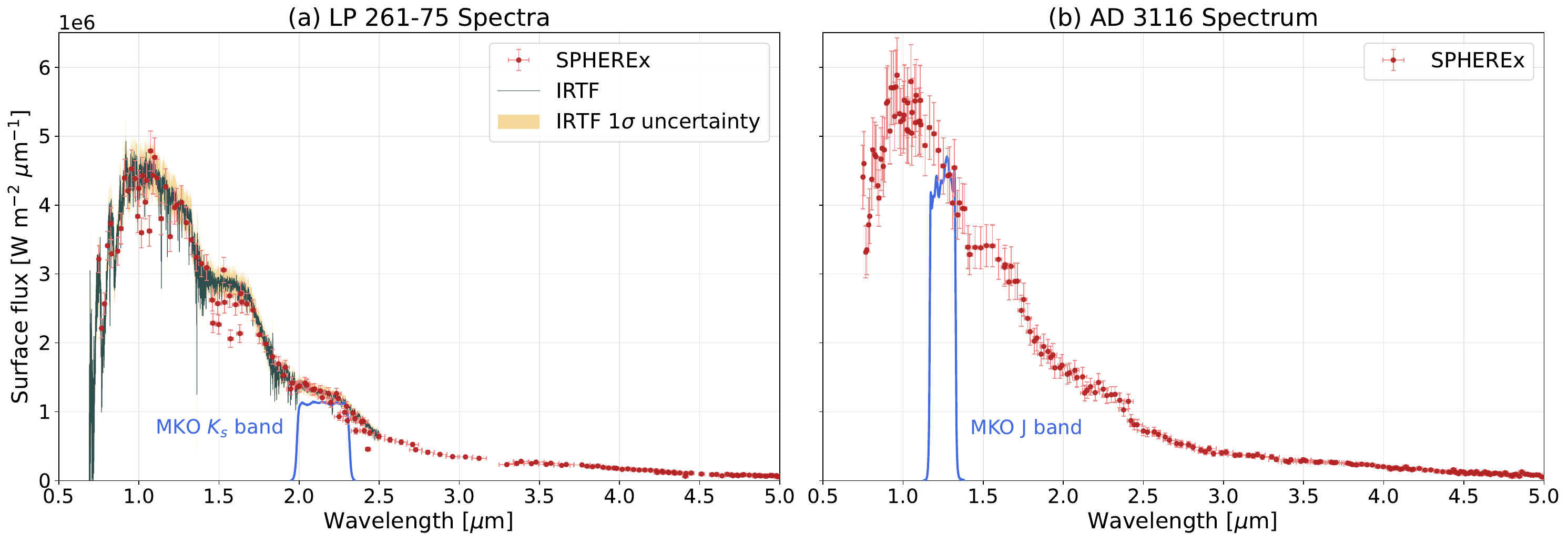}
    \caption{Stellar surface-flux spectra of (a) LP~261-75 and (b) AD~3116. Red points show SPHEREx measurements; vertical error bars denote $1\sigma$ uncertainties while the red horizontal error bars denote the spectral-channel widths. The blue curves show the MKO $K_s$- and $J$-band transmission filters scaled for visibility. In panel (a), the gray-green curve and yellow shaded region show the IRTF spectrum and its $1\sigma$ uncertainty. Both spectra are shown on the same flux scale.}
    \label{fig:stellar_surface_spectra}
\end{figure*}

The stellar spectra are presented in Figure~\ref{fig:stellar_surface_spectra}, with the flux densities expressed at the stellar surface. We converted the flux density observed at Earth, $F_{\lambda,\oplus}$, to the stellar surface flux density according to $F_{\lambda,\star}=fF_{\lambda,\oplus}$, where $f=(d/R_\star)^2$ is the geometric scaling factor, $d$ is the distance to the system, and $R_\star$ is the stellar radius. We determined $f$ through Monte Carlo sampling of Gaussian distributions for the parallax and stellar radius adopted in Tables~\ref{tab1} and~\ref{tab2}. The surface-flux uncertainties include per-channel measurement errors but are dominated by uncertainty in the wavelength-independent scaling factor $f$, corresponding to signal-to-noise ratios of about 16 and 11 for LP~261-75 and AD~3116, respectively. The IRTF/SpeX spectrum of LP~261-75, described in Section~\ref{sec:irtf}, is also included in Figure~\ref{fig:stellar_surface_spectra}.

\subsection{IRTF/SpeX Spectroscopy}\label{sec:irtf}

LP~261-75A was observed with the SpeX spectrograph on the NASA Infrared Telescope Facility (IRTF; \citealt{rayner2003spex}) as part of the COCONUTS survey \citep{zhang2020coconuts, zhang2022coconuts}. The spectrum was collected in SXD mode with a $0\farcs3 \times 15\arcsec$ slit, which spans a wavelength range of $0.7 - 2.55 \, \mu{\rm m}$ with a resolving power of $R \approx 2000$. The data were reduced using the \texttt{Spextool} reduction package \citep{cushing2004spextool}. After applying the same geometric scaling, the SpeX and SPHEREx spectra show good overall agreement, particularly in the $K_s$ band in which we observed the eclipse of LP~261-75C. In contrast, the SpeX spectrum lies marginally above the SPHEREx measurements near $1.5\,\mu{\rm m}$.

Following the procedure outlined by \citet{mann2013metal, mann2014metal}, we used this spectrum to calculate the metallicity of the M dwarf host star. Because LP~261-75A's M4.5~V classification \citep{reid2006lp26175} lies in the overlap of the early- and late-M calibrations, we applied both $K$-band metallicity relations \citep{mann2013metal,mann2014metal} to the rest-frame SpeX spectrum. The \citet{mann2013metal} calibration yielded ${\rm [Fe/H]}=+0.52\pm0.09$ and ${\rm [M/H]}=+0.37\pm0.08$, while the late-type calibration of \citet{mann2014metal} yielded ${\rm [Fe/H]}=+0.59\pm0.08$. These quoted uncertainties combine in quadrature Monte Carlo measurement errors with calibration scatters of 0.08 and 0.07 dex for the early- and late-type relations. Both iron-abundance estimates lie at or slightly above the metal-rich boundary of the calibration samples, which extend to ${\rm [Fe/H]}\simeq+0.56$. We consequently interpret these measurements as evidence that LP 261-75A is metal rich, with ${\rm [Fe/H]}\gtrsim+0.5$, rather than as precise determinations of its metal abundance. The metal richness of LP~261-75A, compared with the near-solar metallicity of the AB Doradus moving group \citep{barenfeld2013abdor, maccarthy2014abdor}, provides additional independent evidence against group membership.

\section{Analysis}\label{sec:analysis}

\subsection{Global Fits}\label{sec:global_fits}

We jointly modeled the space-based photometry, WIRC secondary-eclipse light curves, and radial-velocity measurements described in Section~\ref{sec:observational_data} using the Bayesian modeling framework \texttt{juliet} \citep{espinoza2019}. For each system, the photometric and radial-velocity datasets shared a single set of orbital parameters, allowing the transit shapes, secondary-eclipse timings, and radial-velocity measurements to constrain a self-consistent orbital solution. Within \texttt{juliet}, we used \texttt{batman} for generating the transit and eclipse models \citep{kreidberg2015}, \texttt{radvel} for Keplerian radial velocities \citep{fulton2018}, and \texttt{dynesty} for posterior sampling \citep{speagle2020}. 

We parameterized each orbit using the orbital period \(P\), mid-primary-transit time \(t_0\), eccentricity \(e\), argument of periastron \(\omega\), scaled semimajor axis \(a/R_\star\), and radial-velocity semiamplitude \(K\). We adopted the transit geometry of the \(r_1\)–\(r_2\) parameterization of \citet{julietespinoza2018}, which maps onto the companion-to-star radius ratio \(p=R_{\mathrm{BD}}/R_\star\) and impact parameter \(b\). The TESS and K2 light curves were modeled with combined primary-transit and secondary eclipse models. For these primary transits, we adopted a logarithmic stellar limb-darkening law appropriate for M-dwarf surface-brightness profiles \citep{espinoza2016}. The secondary-eclipse light curves were modeled using \texttt{batman}, which computes the occulted fraction from the projected overlap of the stellar and companion disks, assuming a uniformly bright brown dwarf. To account for the 29.4-min integration time of the K2 observations, we supersampled the K2 light curve model by a factor of 20 during model evaluation. Since the secondary-eclipse phase is $\phi_{\rm sec} \simeq \frac{1}{2} + \frac{2}{\pi}e\cos\omega$ to first order in eccentricity, the WIRC light curve constrained $e\cos\omega$ through the eclipse time predicted by the orbital model.

\begin{figure*}[ht!]
    \centering
    \includegraphics[width=\textwidth]{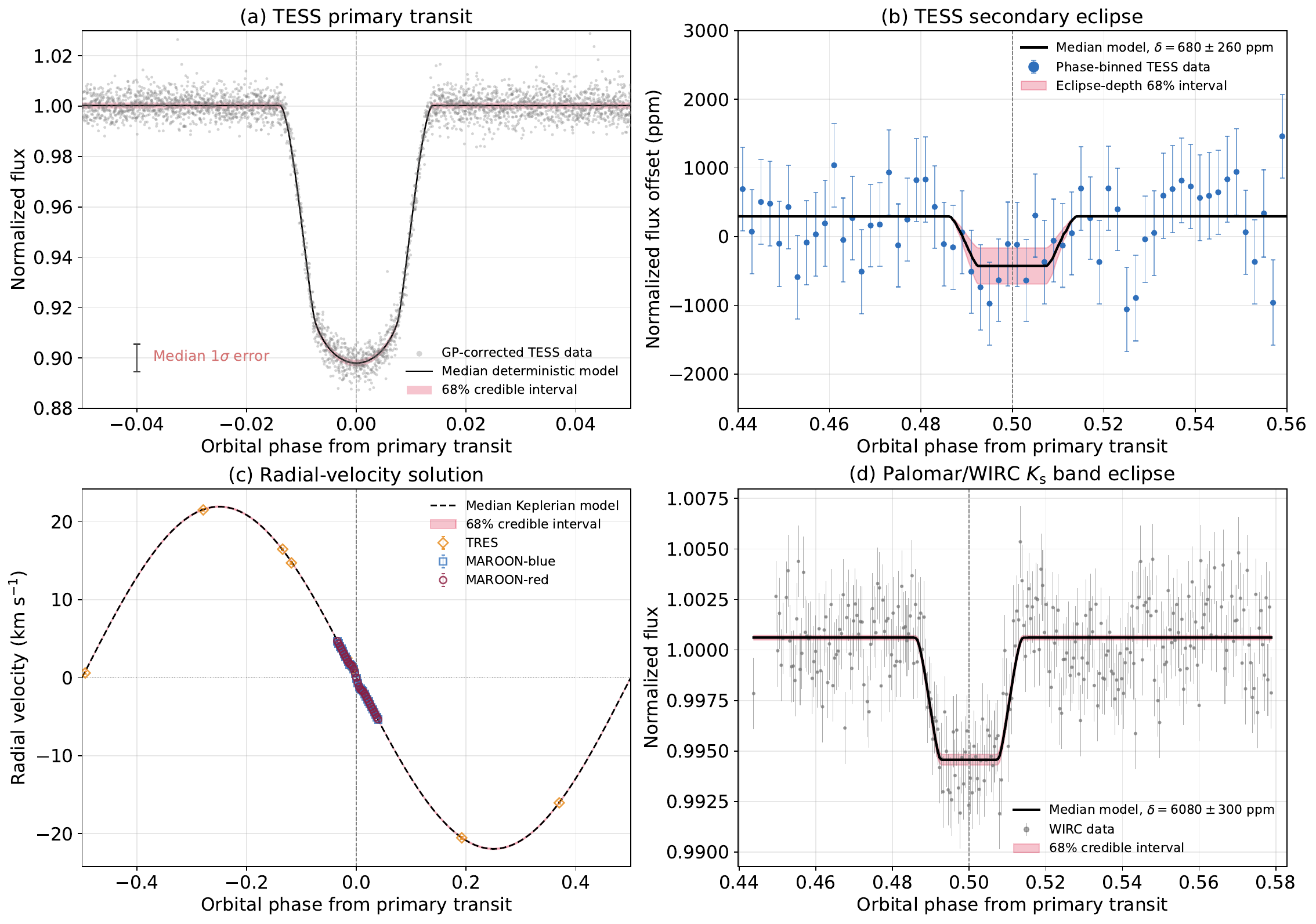}
    \caption{Joint photometric and radial-velocity fit for LP~261-75C. Gray points denote individual photometric measurements, black curves the median models, and red shading the 68\% credible intervals. (a) Phase-folded, GP-corrected TESS primary transit. (b) GP-corrected TESS secondary eclipse; blue points are inverse-variance-weighted orbital-phase bins of width $\Delta\phi=0.002$, with error bars denoting $1\sigma$ uncertainties. The red band represents uncertainty in the eclipse depth with the median out-of-eclipse flux $\mu_{\rm TESS}$ fixed. (c) Phase-folded TRES, MAROON-blue, and MAROON-red radial velocities. The in-transit measurements near phase zero exhibit the Rossiter--McLaughlin effect analyzed by \citet{brady2025maroonx}, which we do not explicitly model. (d) Palomar/WIRC secondary-eclipse photometry with an eclipse depth of $6080 \pm 300$ ppm.}    
    \label{fig:lp261_joint_fit}
\end{figure*}

\begin{figure*}[ht!]
    \centering
    \includegraphics[width=0.95\textwidth]{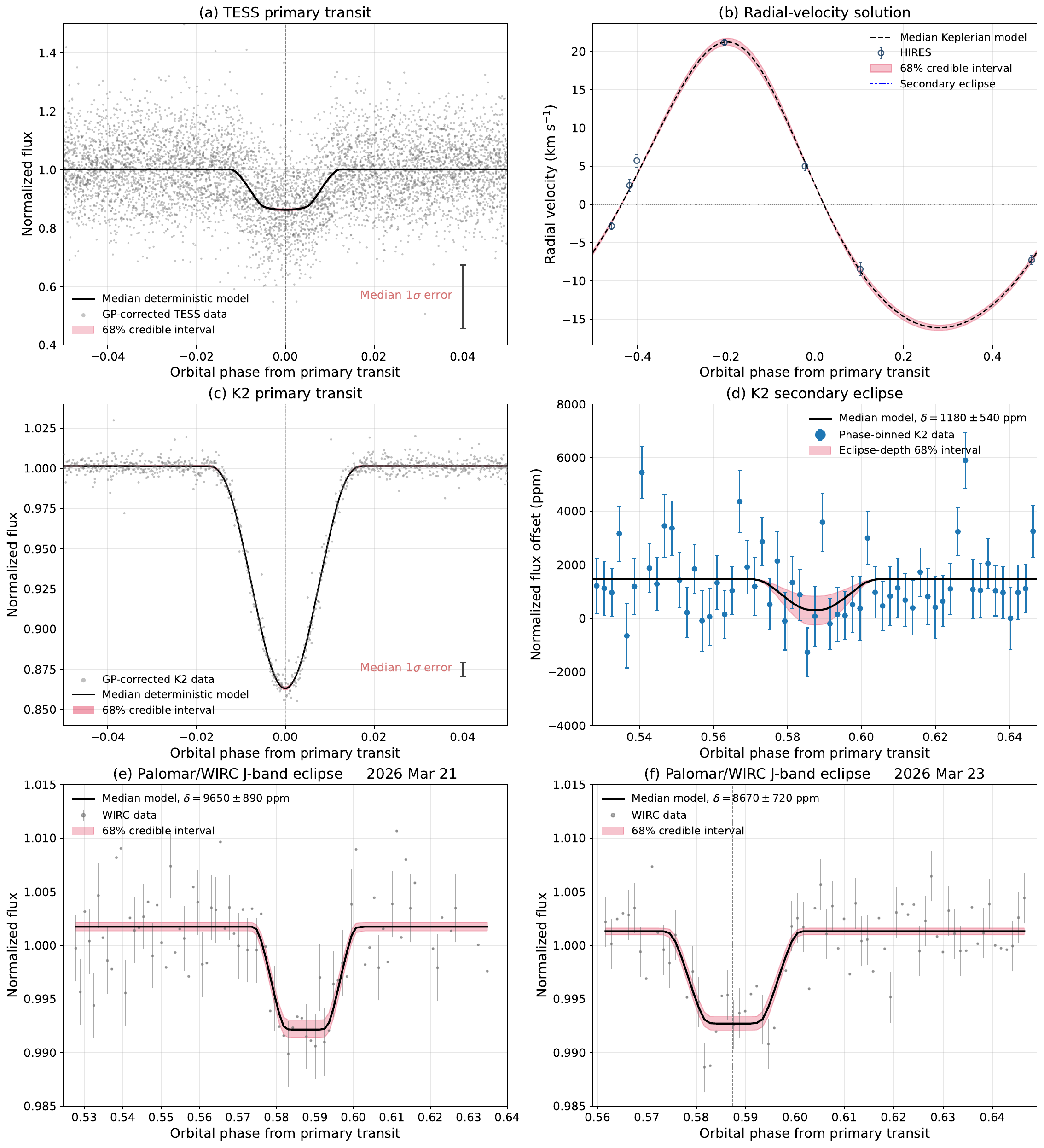}
    \caption{Joint photometric and radial-velocity fit for AD~3116B. Black curves and red shading denote the median models and 68\% credible intervals. Grey points are individual photometric measurements; error bars denote $1\sigma$ uncertainties. (a) Phase-folded, GP-corrected TESS primary transit. The credible interval is narrower than the model line. (b) Phase-folded HIRES radial velocities. Blue points are the HIRES measurements and the blue dashed vertical line marks the secondary-eclipse phase. (c) Phase-folded, GP-corrected K2 primary transit. The comparatively V-shaped profile reflects the 30-minute K2 integration time relative to the $1.2$-hr transit duration. (d) GP-corrected K2 secondary eclipse; blue points are inverse-variance-weighted orbital-phase bins of width $\Delta\phi=0.002$, with error bars denoting $1\sigma$ uncertainties. The red band represents uncertainty in the eclipse depth with the median out-of-eclipse flux $\mu_{\rm K2}$ fixed. (e) and (f) Palomar/WIRC $J$-band secondary eclipses observed on 2026 March 21 and 23, with depths of $9650\pm890$ and $8670\pm720$ ppm, respectively. The secondary-eclipse phase of $\approx0.59$ indicates an eccentric orbit.}
    \label{fig:ad3116b_joint_fit}
\end{figure*}

To derive the companion mass, radius, and semimajor axis, we externally supplied the stellar parameters $M_\star$ and $R_\star$ from version 8 of the TESS Input Catalog (TICv8) as Gaussian distributions, using the values and uncertainties reported in Tables ~\ref{tab1} and \ref{tab2} \citep{dressing2018cooldwarfs, stassun2019tesscatalogue}. These parameters were calculated using empirically calibrated relations based on broadband photometry and Gaia parallaxes, and have been found to be more reliable than parameters derived from stellar models for cool dwarfs \citep[e.g.,][]{dressing2019}. We also note that the TICv8 stellar parameters agree with those derived independently in \cite{irwin2018lp26175} and \citet{gillen2017ad3116} to within $1\sigma$. 

We adopted broad priors on the orbital and instrument nuisance parameters. The uniform priors on $P$ and $t_0$ had half-widths of 0.01 and 0.1 days, respectively, centered on the period and mid-transit time returned by a box least squares (BLS) search. We centered Gaussian priors on $a/R_\star$ using the TIC stellar densities and accounting for the non-negligible companion mass, $M_{\rm{BD}}/M_\star \approx 0.2$. We adopted a conservative width of 1.0 rather than the density-derived uncertainties of 0.1–0.2. For AD 3116B, we additionally restricted $R_{\rm BD}/R_\star$ to $0.2$--$0.5$, following the approach of \citet{gillen2017ad3116}, which imposed $R_{\rm BD}/R_\star<0.46$ to exclude an alternative grazing solution. In our analysis, this grazing solution corresponds to $R_{\rm BD}/R_\star\simeq0.64$ and would imply $R_{\rm BD}\simeq1.85\,R_{\rm J}$. There are only a few transiting brown dwarfs with radii exceeding $1.5\,R_{\rm Jup}$ \citep{carmichael2023bds}, including young systems ($\lesssim150$~Myr) such as RIK-72B and HIP~33609b \citep{david2019, vowell2023}, the strongly irradiated KOI-3886C \citep{lillobox2021}, and the low-mass TOI-4987b, for which prolonged deuterium burning has been proposed \citep{bodenheimer2013, subjak2024}. These examples demonstrate that age and mass alone do not exclude a large brown dwarf radius. However, AD~3116B's older age and greater mass physically disfavour these mechanisms for causing this level of inflation. In Section~\ref{sec:analysis}, we also demonstrate that external radiation and tidal heating are small for this system, making inflation via interaction with the host star highly unlikely. Our adopted priors and resulting posterior constraints are presented in Table~\ref{tab_lp26175c} and Table~\ref{tab_ad3116b}.

\begin{deluxetable*}{lllll}
\tablewidth{0pt}
\tablecaption{Model parameters, priors, and derived quantities for the joint fit of LP~261-75C\label{tab_lp26175c}}
\tablehead{
\colhead{Parameter} & \colhead{Description} & \colhead{Units} & \colhead{Prior} & \colhead{Posterior Median}
}
\startdata
\multicolumn{5}{c}{Shared Model Parameters}\\
\hline
$P$ & Orbital period & days & $\mathcal{U}(1.8715,1.8915)$ & $1.88172274^{+0.00000016}_{-0.00000016}$ \\
$t_{0}$ & Mid-transit time & BJD & $\mathcal{U}(2458870.9236,2458871.1236)$ & $2458871.022447^{+0.000048}_{-0.000046}$ \\
$r_{1}$ & Transit-geometry sampling parameter 1 & --- & $\mathcal{U}(0,1)$ & $0.371^{+0.033}_{-0.025}$ \\
$r_{2}$ & Transit-geometry sampling parameter 2 & --- & $\mathcal{U}(0,1)$ & $0.2933^{+0.0017}_{-0.0019}$ \\
$e$ & Orbital eccentricity & --- & $\mathcal{U}(0,0.05)$ & $0.00073^{+0.00125}_{-0.00053}$ \\
$\omega$ & Argument of periastron & degrees & $\mathcal{U}(0,360)$ & $165^{+103}_{-79}$ \\
$a/R_{\star}$ & Scaled semimajor axis & --- & $\mathcal{N}(14.6,1.0)$ & $14.847^{+0.088}_{-0.097}$ \\
\hline
\multicolumn{5}{c}{Photometric Model Parameters}\\
\hline
$q_{1,\rm TESS}$ & Logarithmic limb-darkening parameter 1 & --- & $\mathcal{U}(0,1)$ & $0.70^{+0.20}_{-0.25}$ \\
$q_{2,\rm TESS}$ & Logarithmic limb-darkening parameter 2 & --- & $\mathcal{U}(0,1)$ & $0.462^{+0.048}_{-0.086}$ \\
$\mu_{\rm TESS}$ & Out-of-transit flux offset & ppm & $\mathcal{N}(0,100000)$ & $-300^{+1200}_{-1200}$ \\
$\sigma_{w,\rm TESS}$ & Photometric jitter & ppm & $\mathcal{U}_{\log}(0.1,10000)$ & $4.7^{+41.0}_{-4.3}$ \\
$f_{p,\rm TESS}$ & Secondary-eclipse depth & ppm & $\mathcal{U}(0,2000)$ & $680^{+260}_{-270}$ \\
$\sigma_{\rm GP,TESS}$ & Gaussian-process amplitude & ppm & $\mathcal{U}_{\log}(1,10^{12})$ & $9820^{+650}_{-620}$ \\
$\rho_{\rm GP,TESS}$ & Gaussian-process timescale & days & $\mathcal{U}_{\log}(0.35,30)$ & $0.3519^{+0.0031}_{-0.0014}$ \\
$\mu_{\rm WIRC}$ & Out-of-transit flux offset & ppm & $\mathcal{U}(-10000,10000)$ & $5460^{+270}_{-280}$ \\
$\sigma_{w,\rm WIRC}$ & Photometric jitter & ppm & $\mathcal{U}_{\log}(1,10000)$ & $36^{+236}_{-33}$ \\
$f_{p,\rm WIRC}$ & Secondary-eclipse depth & ppm & $\mathcal{U}(1000,20000)$ & $6080^{+300}_{-310}$ \\
\hline
\multicolumn{5}{c}{RV Model Parameters}\\
\hline
$K$ & Radial-velocity semiamplitude & m s$^{-1}$ & $\mathcal{U}(0,50000)$ & $21939^{+76}_{-77}$ \\
$\mu_{\rm red}$ & Systemic radial velocity & m s$^{-1}$ & $\mathcal{U}(-50000,50000)$ & $1526^{+36}_{-35}$ \\
$\mu_{\rm blue}$ & Systemic radial velocity & m s$^{-1}$ & $\mathcal{U}(-50000,50000)$ & $1742^{+34}_{-35}$ \\
$\mu_{\rm TRES}$ & Systemic radial velocity & m s$^{-1}$ & $\mathcal{U}(-50000,50000)$ & $-5196^{+57}_{-54}$ \\
$\sigma_{w,\rm red}$ & Radial-velocity jitter & m s$^{-1}$ & $\mathcal{U}_{\log}(0.001,5000)$ & $200^{+30}_{-24}$ \\
$\sigma_{w,\rm blue}$ & Radial-velocity jitter & m s$^{-1}$ & $\mathcal{U}_{\log}(0.001,5000)$ & $195^{+28}_{-22}$ \\
$\sigma_{w,\rm TRES}$ & Radial-velocity jitter & m s$^{-1}$ & $\mathcal{U}_{\log}(0.001,5000)$ & $15^{+70}_{-14}$ \\
\hline
\multicolumn{5}{c}{Derived Quantities}\\
\hline
$e\cos{\omega}$ & Eccentricity vector component & --- & --- & $-0.00012^{+0.00030}_{-0.00043}$ \\
$e\sin{\omega}$ & Eccentricity vector component & --- & --- & $0.0000^{+0.0011}_{-0.0010}$ \\
$i$ & Orbital inclination & degrees & --- & $89.78^{+0.15}_{-0.19}$ \\
$M_{2}$ & Secondary mass & $M_{\rm Jup}$ & --- & $65.7^{+2.9}_{-2.9}$ \\
$R_{2}$ & Secondary radius & $R_{\rm Jup}$ & --- & $0.874^{+0.026}_{-0.027}$ \\
$a_{\mathrm{transit}}$ & Semimajor axis inferred from fitted $a/R_{\star}$ & AU & --- & 
$0.02115^{+0.00064}_{-0.00065}$ \\
$a_{\mathrm{kepler}}$ & Semimajor axis inferred from Kepler's third law & AU & --- & $0.02095^{+0.00045}_{-0.00047}$ \\
$\log{g_{2}}$ & Secondary surface gravity & $\log_{10}({\rm cm\,s^{-2}})$ & --- & $5.329^{+0.033}_{-0.033}$ \\
$\rho_{2}$ & Secondary bulk density & g cm$^{-3}$ & --- & $122^{+13}_{-12}$ \\
$\phi_{\rm sec}$ & Secondary-eclipse phase & --- & --- & $0.49992^{+0.00019}_{-0.00028}$ \\
$t_{\rm sec}$ & Mid-secondary-eclipse time & BJD & --- & $2461052.87980^{+0.00039}_{-0.00048}$ \\
$R_{2}/R_{\star}$ & Secondary-to-primary radius ratio & --- & --- & $0.2933^{+0.0017}_{-0.0019}$ \\
$b$ & Transit impact parameter & --- & --- & $0.056^{+0.049}_{-0.038}$ \\
\enddata
\tablecomments{Posterior values and uncertainties are the 50th, 16th, and 84th percentiles of the posterior distributions. For the priors, $\mathcal{U}(l,u)$, $\mathcal{N}(\mu,\sigma)$, and $\mathcal{U}_{\log}(l,u)$ denote uniform, normal, and log-uniform distributions. The parameters $r_{1}$ and $r_{2}$ are the Espinoza transit-geometry sampling parameters that map to $b$ and $R_{2}/R_{\star}$. The parameters $\mu_{\rm TESS}$ and $\mu_{\rm WIRC}$ describe the mean out-of-transit flux offsets. The transit-derived semimajor axis is calculated from the fitted $a/R_{\star}$ and adopted stellar radius, whereas the Keplerian semimajor axis is calculated from the orbital period and total system mass.}
\end{deluxetable*}
\begin{deluxetable*}{lllll}
\tablewidth{0pt}
\tablecaption{Model parameters, priors, and derived quantities for the joint fit of AD~3116B\label{tab_ad3116b}}
\tablehead{
\colhead{Parameter} & \colhead{Description} & \colhead{Units} & \colhead{Prior} & \colhead{Posterior Median}
}
\startdata
\multicolumn{5}{c}{Shared Model Parameters}\\
\hline
$P$ & Orbital period & days & $\mathcal{U}(1.9729659319,1.9929659319)$ & $1.98279424^{+0.00000021}_{-0.00000021}$ \\
$t_{0}$ & Mid-transit time & BJD & $\mathcal{U}(2459500.568881,2459500.768881)$ & $2459500.66989^{+0.00019}_{-0.00019}$ \\
$r_{1}$ & Transit-geometry sampling parameter 1 & --- & $\mathcal{U}(0,1)$ & $0.373^{+0.044}_{-0.048}$ \\
$r_{2}$ & Transit-geometry sampling parameter 2 & --- & $\mathcal{U}(0,1)$ & $0.504^{+0.022}_{-0.029}$ \\
$e$ & Orbital eccentricity & --- & $\mathcal{U}(0,0.3)$ & $0.1421^{+0.0058}_{-0.0032}$ \\
$\omega$ & Argument of periastron & degrees & $\mathcal{U}(-180,180)$ & $14.3^{+7.3}_{-7.2}$ \\
$a/R_{\star}$ & Scaled semimajor axis & --- & $\mathcal{N}(15.55,1.0)$ & $16.15^{+0.40}_{-0.44}$ \\
\hline
\multicolumn{5}{c}{Photometric Model Parameters}\\
\hline
$q_{1,\mathrm{TESS}}$ & Logarithmic limb-darkening parameter 1 & --- & $\mathcal{U}(0,1)$ & $0.43^{+0.36}_{-0.29}$ \\
$q_{2,\mathrm{TESS}}$ & Logarithmic limb-darkening parameter 2 & --- & $\mathcal{U}(0,1)$ & $0.53^{+0.27}_{-0.26}$ \\
$\mu_{\mathrm{TESS}}$ & Out-of-transit flux offset & ppm & $\mathcal{N}(0,200000)$ & $-1160^{+640}_{-650}$ \\
$\sigma_{w,\mathrm{TESS}}$ & Photometric jitter & ppm & $\mathcal{U}_{\log}(0.1,100000)$ & $16^{+322}_{-15}$ \\
$f_{p,\mathrm{TESS}}$ & Secondary-eclipse depth & ppm & $\mathcal{U}(0,20000)$ & $6000^{+3200}_{-3100}$ \\
$\sigma_{\mathrm{GP,TESS}}$ & Gaussian-process amplitude & ppm & $\mathcal{U}_{\log}(1,10^{12})$ & $4810^{+810}_{-820}$ \\
$\rho_{\mathrm{GP,TESS}}$ & Gaussian-process timescale & days & $\mathcal{U}_{\log}(0.35,30)$ & $0.398^{+0.079}_{-0.035}$ \\
$q_{1,\mathrm{K2}}$ & Logarithmic limb-darkening parameter 1 & --- & $\mathcal{U}(0,1)$ & $0.43^{+0.35}_{-0.28}$ \\
$q_{2,\mathrm{K2}}$ & Logarithmic limb-darkening parameter 2 & --- & $\mathcal{U}(0,1)$ & $0.31^{+0.22}_{-0.19}$ \\
$\mu_{\mathrm{K2}}$ & Out-of-transit flux offset & ppm & $\mathcal{U}(-100000,100000)$ & $-1480^{+800}_{-840}$ \\
$\sigma_{w,\mathrm{K2}}$ & Photometric jitter & ppm & $\mathcal{U}_{\log}(0.1,100000)$ & $2956^{+81}_{-78}$ \\
$f_{p,\mathrm{K2}}$ & Secondary-eclipse depth & ppm & $\mathcal{U}(0,20000)$ & $1180^{+550}_{-520}$ \\
$\sigma_{\mathrm{GP,K2}}$ & Gaussian-process amplitude & ppm & $\mathcal{U}_{\log}(1,10^{12})$ & $13000^{+420}_{-390}$ \\
$\rho_{\mathrm{GP,K2}}$ & Gaussian-process timescale & days & $\mathcal{U}_{\log}(0.35,30)$ & $0.372^{+0.018}_{-0.014}$ \\
$\mu_{\mathrm{WIRC1}}$ & Out-of-transit flux offset & ppm & $\mathcal{U}(-13000,13000)$ & $7880^{+800}_{-850}$ \\
$\sigma_{w,\mathrm{WIRC1}}$ & Photometric jitter & ppm & $\mathcal{U}_{\log}(1,10000)$ & $38^{+465}_{-34}$ \\
$f_{p,\mathrm{WIRC1}}$ & Secondary-eclipse depth & ppm & $\mathcal{U}(1000,15000)$ & $9650^{+880}_{-900}$ \\
$\mu_{\mathrm{WIRC2}}$ & Out-of-transit flux offset & ppm & $\mathcal{U}(-13000,13000)$ & $7380^{+640}_{-650}$ \\
$\sigma_{w,\mathrm{WIRC2}}$ & Photometric jitter & ppm & $\mathcal{U}_{\log}(1,10000)$ & $53^{+501}_{-49}$ \\
$f_{p,\mathrm{WIRC2}}$ & Secondary-eclipse depth & ppm & $\mathcal{U}(1000,15000)$ & $8670^{+740}_{-700}$ \\
\hline
\multicolumn{5}{c}{RV Model Parameters}\\
\hline
$K$ & Radial-velocity semiamplitude & m s$^{-1}$ & $\mathcal{U}(0,50000)$ & $18730^{+410}_{-400}$ \\
$\mu_{\mathrm{HIRES}}$ & Systemic radial velocity & m s$^{-1}$ & $\mathcal{U}(-50000,50000)$ & $34730^{+280}_{-280}$ \\
$\sigma_{w,\mathrm{HIRES}}$ & Radial-velocity jitter & m s$^{-1}$ & $\mathcal{U}_{\log}(0.001,10000)$ & $150^{+520}_{-140}$ \\
\hline
\multicolumn{5}{c}{Derived Quantities}\\
\hline
$e\cos{\omega}$ & Eccentricity vector component & --- & --- & $0.13756^{+0.00069}_{-0.00071}$ \\
$e\sin{\omega}$ & Eccentricity vector component & --- & --- & $0.035^{+0.019}_{-0.018}$ \\
$i$ & Orbital inclination & degrees & --- & $88.67^{+0.26}_{-0.26}$ \\
$M_{2}$ & Secondary mass & $M_{\mathrm{Jup}}$ & --- & $53.6^{+3.1}_{-3.1}$ \\
$R_{2}$ & Secondary radius & $R_{\mathrm{Jup}}$ & --- & $0.999^{+0.045}_{-0.045}$ \\
$a_{\mathrm{transit}}$ & Semimajor axis inferred from fitted $a/R_{\star}$ & AU & --- & $0.0220^{+0.0010}_{-0.0010}$ \\
$a_{\mathrm{kepler}}$ & Semimajor axis inferred from Kepler's third law & AU & --- & $0.02112^{+0.00053}_{-0.00056}$ \\
$\log{g_{2}}$ & Secondary surface gravity & $\log_{10}({\rm cm\,s^{-2}})$ & --- & $5.124^{+0.046}_{-0.046}$ \\
$\rho_{2}$ & Secondary bulk density & g cm$^{-3}$ & --- & $66.6^{+10.5}_{-8.9}$ \\
$\phi_{\mathrm{sec}}$ & Secondary-eclipse phase & --- & --- & $0.58737^{+0.00043}_{-0.00044}$ \\
$t_{\mathrm{sec}}$ & Mid-secondary-eclipse time & BJD & --- & $2461121.77742^{+0.00077}_{-0.00082}$ \\
$R_{2}/R_{\star}$ & Secondary-to-primary radius ratio & --- & --- & $0.3511^{+0.0067}_{-0.0086}$ \\
$b$ & Transit impact parameter & --- & --- & $0.353^{+0.060}_{-0.064}$ \\
\enddata
\tablecomments{This table adopts the same conventions as Table~\ref{tab_lp26175c} for posterior percentiles, prior notation, and parameter definitions. For the $r_{1}$ and $r_{2}$ parameterization, the radius-ratio bounds are $p_{l}=0.2$ and $p_{u}=0.5$.}
\end{deluxetable*}

For the final \texttt{dynesty} fits, we used $N_{\rm live}=600$ with \texttt{rwalk} sampling for LP 261-75C and $N_{\rm live}=800$ with \texttt{rslice} sampling for AD 3116B. The larger number of live points reflects the greater dimensionality introduced by the K2 data and second WIRC observing night. For this system, \texttt{rslice} more efficiently explored the posterior ridge in the $(e\cos\omega,e\sin\omega)$ plane associated with AD~3116B's eccentric orbit. The corresponding primary-transit, secondary-eclipse, and radial-velocity fits are shown in Figures~\ref{fig:lp261_joint_fit}--\ref{fig:ad3116b_joint_fit}. As a consistency check, we computed Lomb-Scargle periodograms on the inferred GP component. LP 261-75A and AD 3116A exhibited dominant peaks of \(P \simeq 2.22\,\mathrm{d}\) and \(P \simeq 2.25\,\mathrm{d}\), respectively, consistent with the previously reported rotation periods \citep{gillen2017ad3116, brady2025maroonx}.

For LP~261-75C, the joint fit yields a brown-dwarf mass of $M_{\mathrm{BD}}=65.7 \pm 2.9\,M_{\mathrm{Jup}}$ and a radius of $R_{\mathrm{BD}}=0.874\pm0.027\,R_{\mathrm{Jup}}$ on a short-period orbit with $P\simeq1.88\,\mathrm{d}$ and $a\simeq0.02115\,\mathrm{AU}$. These values are consistent with the system parameters reported by \citet{brady2025maroonx}. The primary-transit fit shown in panel (a) of Figure~\ref{fig:lp261_joint_fit} constrains the radius ratio to $R_{\mathrm{BD}}/R_\star=0.2933^{+0.0017}_{-0.0019}$, which enters directly into the eclipse depth calculation in Section~\ref{sec:internal_temp}. The orbit is nearly circular, with $e=0.00073^{+0.00125}_{-0.00053}$, a one-sided $2\sigma$ upper limit of $e<0.0042$, and a secondary-eclipse phase of $\phi_{\mathrm{sec}}=0.49992^{+0.00019}_{-0.00028}$. The previously available transit and radial-velocity data already required a nearly circular orbit for LP~261-75C, meaning the secondary-eclipse timing minimally influenced the orbit geometry (in contrast to the appreciably eccentric AD~3116B system discussed below). The TESS photometry features a $\sim 2.5\sigma$ secondary-eclipse posterior with a depth of $680^{+260}_{-270}$ ppm and a $3\sigma$ upper limit of $1410$ ppm, shown in panel (b). Although the TESS measurements binned at intervals of $\Delta\phi=0.002$ exhibit considerable scatter, they show a slight flux decrease at the expected phase. The in-transit MAROON-X radial velocities in panel (c) also exhibit the Rossiter--McLaughlin effect previously analyzed by \citet{brady2025maroonx}. Finally, the WIRC $K_s$-band photometry in panel (d) provides a clear secondary-eclipse measurement, with a depth of $6080\pm300\,\mathrm{ppm}$. This near-infrared eclipse depth, together with the mass and radius-ratio constraints, provides the key observational input to our intrinsic-temperature inference in Section~\ref{sec:internal_temp}.

For AD~3116B, the joint fit yields a brown-dwarf mass of $M_{\mathrm{BD}}=53.6\pm 3.1\,M_{\mathrm{Jup}}$ and a radius of $R_{\mathrm{BD}}=0.999\pm0.045\,R_{\mathrm{Jup}}$ on a similarly short-period orbit of $P\simeq1.98\,\mathrm{d}$ and a semimajor axis of $a\simeq0.0220\,\mathrm{AU}$, consistent with previous determinations \citep{gillen2017ad3116,carmichael2023bds}. The joint TESS and K2 transit fits shown in panels (a) and (c) of Figure~\ref{fig:ad3116b_joint_fit} constrain the shared radius ratio to $R_{\mathrm{BD}}/R_\star=0.3511^{+0.0067}_{-0.0086}$.  We obtain weakly constrained TESS- and K2-band eclipse depths of $6000^{+3200}_{-3100}$ ppm and $1180^{+550}_{-520}$ ppm, respectively, with $3\sigma$ upper limits of $15{,}500$ ppm and $2640$ ppm. Although the phase-folded light curve in panel (d) shows a slight K2 flux decrease, its eclipse depth is consistent with the photometric noise. The comparatively large scatter in the TESS photometry is unsurprising because AD~3116 is approximately 5.5 times more distant and thus 3.6 magnitudes fainter in the TESS band than LP~261-75, despite the comparable luminosities of the two M dwarfs.

The eccentricity of AD~3116B is evident from both the secondary-eclipse phase of $\phi_{\mathrm{sec}}=0.58737$ and the non-sinusoidal radial-velocity curve shown in panel (b). The joint fit yields $e=0.1421^{+0.0058}_{-0.0032}$ and $\omega=14.3^{+7.3}_{-7.2}\,\mathrm{deg}$, corresponding to $e\cos\omega=0.13756^{+0.00069}_{-0.00071}$ and $e\sin\omega=0.035^{+0.019}_{-0.018}$. The eclipse timing therefore provides a particularly strong constraint on $e\cos\omega$, whereas $e\sin\omega$ remains less precisely determined. For comparison, \citet{carmichael2023bds} determined $e\cos\omega=0.128^{+0.028}_{-0.025}$, $e\sin\omega=0.059^{+0.055}_{-0.048}$, and $e=0.144^{+0.045}_{-0.025}$ without secondary-eclipse observations, indicating our new fit tightly constrains eccentricity. The companion mass is inferred from the radial-velocity semiamplitude, fitted orbital parameters, and adopted stellar mass through 
\begingroup 
\setlength{\abovedisplayskip}{2pt}
\setlength{\abovedisplayshortskip}{1pt}
\begin{equation}
K = \left(\frac{2\pi G}{P}\right)^{1/3}
\frac{M_{\mathrm{BD}}\sin i}{\left(M_\star + M_{\mathrm{BD}}\right)^{2/3}}
\frac{1}{\sqrt{1-e^2}}.
\label{eq:rv_semiamplitude}
\end{equation}
\endgroup
By tightly constraining $e\cos\omega$, the covariance amongst $e$, $\omega$, and $K$ is also reduced in the joint fit. We thus find $M_{\mathrm{BD}}=53.6\pm 3.1\,M_{\mathrm{Jup}}$, improving the mass precision by a factor of two relative to $54.6^{+6.8}_{-6.9}\,M_{\mathrm{Jup}}$ measured in \citet{carmichael2023bds}. This improvement is important for our evolutionary analysis because brown-dwarf cooling models map mass and age onto intrinsic temperature. The two WIRC observations yield $J$-band secondary-eclipse depths of $9650^{+880}_{-900}$ ppm and $8670^{+740}_{-700}$ ppm, as shown in panels (e) and (f) of Figure~\ref{fig:ad3116b_joint_fit}. We fitted the two nights with independent eclipse depths to allow for possible brown-dwarf variability on day-long timescales; however, the two measurements differ by only $0.9\sigma$. The two observing nights yield an inverse-variance weighted eclipse depth of $9060\pm560$ ppm. Taken together, the global fits provide the companion masses, radius ratios, and near-infrared eclipse depths required by the intrinsic-temperature analysis in Section~\ref{sec:internal_temp}. We use these measurements alongside brown dwarf evolutionary models to infer the two brown dwarf intrinsic temperatures and, by inverting the mass--temperature--age relation, estimate the ages of their host systems.


\subsection{Background on Evolutionary Models} \label{sec:Evolutionary Models}


Successive generations of brown dwarf evolutionary models have sought to reproduce the observable properties of their substellar populations \citep[e.g.,][]{burrows1997,baraffe2003,saumon2008,phillips2020,marley2021}. Unable to sustain hydrogen fusion, brown dwarfs evolve primarily through cooling and contraction, radiating residual energy from their gravitational collapse and formation. During their early evolution, brown dwarfs with masses $M \gtrsim 13\,M_{\mathrm{Jup}}$ reach central temperatures sufficient to ignite deuterium, temporarily slowing their cooling \citep{spiegel2011}. For an assumed composition and atmosphere treatment, evolutionary models predict the bolometric luminosity, radius, and effective temperature as functions of mass and age. These models are widely used to characterize free-floating brown dwarfs in stellar clusters and the local field. For example, a measured bolometric luminosity and an age estimated from cluster or moving-group membership can be combined to infer a model-dependent mass \citep[e.g.,][]{filippazzo2015,faherty2016,zapatero2018}. Systems with independently constrained masses, ages, and luminosities therefore provide tests of evolutionary models.

Early non-grey atmosphere and evolutionary calculations reproduced key molecular features in the spectrum of Gliese 229B and used this system's spectrum and luminosity to constrain its physical properties \citep{allard1996,marley1996}. \citet{burrows1997} subsequently presented a broader grid of coupled evolutionary and spectral models spanning giant planets and brown dwarfs. Schematically, these models solve for the interior structure, apply the atmospheric boundary condition to determine luminosity, and update the entropy through energy conservation, as described below. This cycle is repeated and governs a brown dwarf's evolution.

A common approach begins with a chosen mass $M$, elemental composition, and initial specific entropy $s_0$. Under the assumption of efficient convection, the brown dwarf’s interior is approximately uniform in composition and specific entropy. A trial central pressure is selected, and a precomputed equation of state (EOS) supplies the corresponding central density and temperature for the adopted composition and entropy. The spherically symmetric model is then integrated outward using mass conservation and hydrostatic equilibrium. These equations determine how the enclosed mass increases and the pressure decreases outward. At each new radius $r+\Delta r$, the EOS is further evaluated using the updated pressure and the same entropy,
\begin{equation}
\operatorname{EOS}\!\bigl[P(r+\Delta r),s(t)\bigr]
=
\bigl[T(r+\Delta r),\rho(r+\Delta r)\bigr].
\label{eq:eos_update}
\end{equation}
This entropy is initialized as $s(0)=s_0$, held fixed during each interior integration, and updated when advancing the evolution in time. The central pressure is adjusted and the calculation is repeated until the model satisfies the total-mass condition $m(R)=M$, where $R$ is the outer radius. The resulting solution gives the brown dwarf's radius and its internal pressure, density, and temperature profiles.

To provide the outer boundary condition for the evolutionary calculation, atmospheric structures and spectra are calculated for a grid of surface gravities $\log g$ and intrinsic temperatures $T_{\mathrm{int}}$. The intrinsic temperature parametrizes the net outward energy flux supplied by the interior. In the absence of external irradiation, this flux equals the wavelength-integrated emergent surface flux,
\begin{equation}
F_{\mathrm{int}}
= \int_0^\infty F_\lambda\,d\lambda
= \sigma_{\mathrm{SB}}T_{\mathrm{int}}^4,
\label{eq:intrinsic_flux}
\end{equation}
where $F_\lambda$ is the emergent spectral flux density and $\sigma_{\mathrm{SB}}$ is the Stefan--Boltzmann constant. Likewise, in the deep, optically thick atmosphere, the model imposes an intrinsic flux $F_{\mathrm{int}}$ to represent the energy supplied by the interior. As this radiation propagates through the atmosphere, wavelength-dependent molecular and cloud opacities reshape the emergent spectrum and produce its characteristic absorption and emission features. To produce the spectra, a pressure–temperature profile is iteratively adjusted until the atmosphere reaches radiative–convective equilibrium. Radiative transfer is solved through discretized atmospheric layers, accounting for local thermal emission, absorption, and scattering.

To apply the atmospheric boundary condition, the surface gravity is calculated from the interior solution as $g=GM/R^2$. The EOS is then used to evaluate the entropy at the convective base of each atmosphere from its local pressure and temperature. At the required gravity, the calculation identifies the atmosphere profile whose entropy at the convective base matches the interior profile entropy, determining the corresponding $T_{\mathrm{int}}$ and hence the luminosity, $L=4\pi R^2\sigma_{\mathrm{SB}}T_{\mathrm{int}}^4$. This luminosity is then used to advance the entropy in time through energy conservation \citep{saumon2008},
\begin{equation}
\left(L-\int_0^M \epsilon_{\mathrm{nuc}}\,dm\right)dt
= -\int_0^M T\,ds\,dm,
\label{eq:entropy_evolution}
\end{equation}
where $\epsilon_{\mathrm{nuc}}$ is the nuclear energy-generation rate per unit mass, $T$ is the local interior temperature, and $ds$ is the change in specific entropy of a fixed mass shell over the time interval $dt$. For a nearly isentropic interior, $ds$ is approximately uniform throughout and can be taken outside the integral, allowing the energy-conservation equation to be solved for $ds/dt$. Over a time step $\Delta t$, the entropy is updated as $s(t+\Delta t)=s(t)+\Delta s$, and the interior structure and atmospheric match are recalculated. Repeating this process forms the basis of brown dwarf cooling calculations, yielding the radius, luminosity, effective temperature, and emergent spectrum as functions of age. For specific details on the physical assumptions and numerical implementation of these models, we refer to \citet{chabrier1997,saumon2008,phillips2020}.

\subsection{Evolutionary Model Evaluation} \label{sec:evolutionary_models_evaluation}

In this work, we compare five sets of evolutionary models to infer the effective temperatures and ages of our two brown dwarf systems: ATMO 2020 \citep{phillips2020}, the updated calculations of \citet{chabrier2003} (hereafter CBPD), Sonora Bobcat \citep{marley2021}, Sonora Diamondback \citep{morley2024diamondback}, and the models of \citet{saumon2008} (hereafter SM08). These models differ in their interior equations of state, atmospheric opacity data, elemental abundances, and treatment of condensate clouds, allowing us to examine how these assumptions affect our method of inferring $\mathrm{T_{int}}$ and age. 

We adopt solar-metallicity, chemical-equilibrium ATMO 2020 models as a cloudless baseline. These models include elemental depletion through condensation and rainout, but do not feature cloud opacity in their radiative transfer. The evolutionary calculations use the H/He EOS of \citet{chabrier2019}. The newer CBPD retains the ATMO atmospheric models but replaces the earlier H/He EOS with the updated version of \citet{chabrierdebras2021}. Comparing these grids therefore isolates the effect of the EOS on cooling and the inferred age. Sonora Bobcat provides independent cloudless, chemical-equilibrium models with different molecular and alkali opacity data and the \citet{saumon1995} H/He EOS. Its evolutionary sequences span $\mathrm{[M/H]}=-0.5,0.0,+0.5$, allowing us to assess the effects of metallicity.

SM08 and Sonora Diamondback use cloudy, chemical-equilibrium atmospheres as evolutionary boundary conditions. Both adopt the \citet{ackerman2001} prescription, in which the factor $f_{\mathrm{sed}}$ controls the balance between upward mixing and condensate settling. Cloud opacity changes the atmospheric temperature structure and therefore the boundary condition that regulates interior cooling. The more recent Diamondback updates the gas opacities, cloud condensation, and mixing treatments, and includes $\mathrm{MgSiO_3}$, $\mathrm{Mg_2SiO_4}$, $\mathrm{Al_2O_3}$, and $\mathrm{Fe}$ clouds. It also adopts the \citet{chabrier2019} H/He EOS, explicitly includes interior heavy elements, and spans $\mathrm{[M/H]}=-0.5,0.0,+0.5$. 

Both model families provide ``hybrid" cooling sequences that connect cloudy and cloudless atmospheric boundary conditions across the L/T transition. SM08 linearly interpolates between cloudy ($f_{\mathrm{sed}}=2$) boundary conditions at $T_{\mathrm{eff}}=1400\,\mathrm{K}$ and cloudless boundary conditions at $1200\,\mathrm{K}$. Diamondback instead adopts cloudy ($f_{\mathrm{sed}}=2$) boundary spectra above $1300\,\mathrm{K}$ and cloudless boundary conditions below. Its hybrid-gravity sequence adopts an empirically motivated gravity dependence for the cloud-clearing temperature. Linear interpolation of this empirical relation gives cloud-clearing temperatures of $\approx1360\,\mathrm{K}$ and $\approx1320\,\mathrm{K}$ for LP 261-75C and AD 3116B surface gravities. 

Both of our systems yield median inferred temperatures above $1400 \, \rm{K}$, so we expect the L/T cloud transition to have only a small effect on our age inference. However, $13.2\%$ of the SM08 temperature posterior for LP~261-75C falls below this threshold, within the cloud-clearing transition. Because the adopted SM08 grid provides no spectra for partial cloud clearing, we pair cloudy $f_{\mathrm{sed}}=2$ spectra with hybrid cooling tracks, retaining the cloudy spectra for this posterior tail as an approximation rather than interpolating between intermediate cloud states.

\begin{figure}[ht!]
    \centering
    \includegraphics[width=\columnwidth]{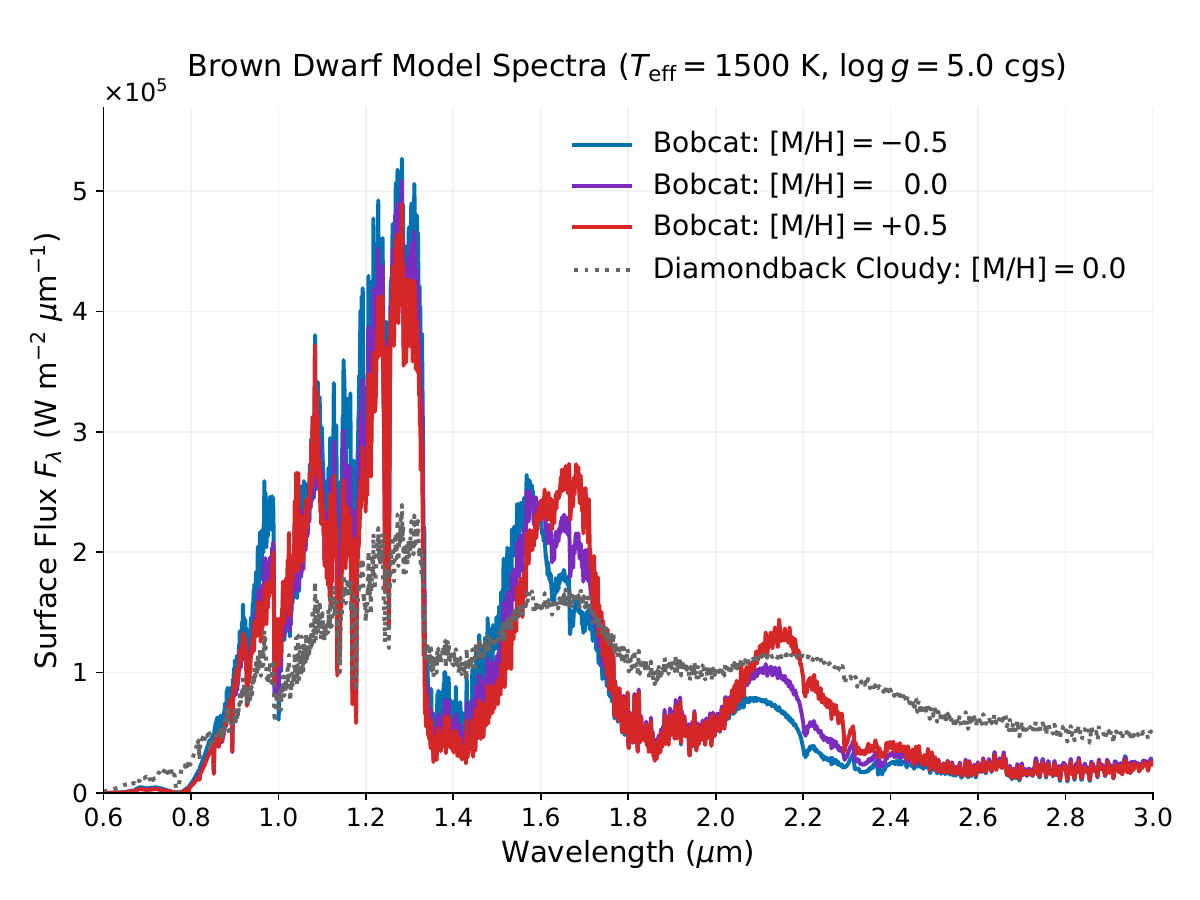}
    \caption{Sonora model spectra at $T_{\mathrm{eff}} = 1500$ K and $\log g = 5.0$. Blue, purple, and red solid curves show cloud-free Bobcat models with $[\mathrm{M/H}] = -0.5$, $0.0$, and $+0.5$, respectively. The grey dotted curve shows the solar-metallicity Diamondback cloudy model with $f_{\mathrm{sed}} = 2$. All spectra are smoothed with a Gaussian kernel ($R = 1200$).}
    \label{fig:metal_cloud_spectra}
\end{figure}

Figure~\ref{fig:metal_cloud_spectra} illustrates the effects of metallicity and clouds at fixed effective temperature and surface gravity by comparing Sonora Bobcat spectra at three metallicities alongside a cloudy, solar-metallicity Sonora Diamondback spectrum. At $T_{\mathrm{int}}=1500\,\mathrm{K}$ and $\log g=5.0$, the Sonora Bobcat model with $\mathrm{[M/H]}=-0.5$ produces stronger $J$-band emission, causing a lower inferred intrinsic temperature for the same $J$-band eclipse depth and radius. Conversely, the $\mathrm{[M/H]}=+0.5$ model produces stronger emission near $1.7\,\mu\mathrm{m}$ in the $H$ band and $2.1\,\mu\mathrm{m}$ in the $K$ band, with the latter approximately $1.5$ times brighter than the $\mathrm{[M/H]}=-0.5$ model. The solar-metallicity, cloudy Diamondback model with $f_{\mathrm{sed}}=2$ has weaker $J$- and $H$-band peaks than the cloudless Bobcat models, a comparable $K$-band peak, and enhanced emission at longer wavelengths. These spectral differences affect the inferred intrinsic temperature, but the resulting age shifts also depend on the corresponding evolutionary tracks, which respond to changes in atmospheric and interior assumptions.

To limit interpolation effects when propagating our mass posteriors in the range of $50$–$70\,M_{\mathrm{Jup}}$, we selected model grids with mass spacings of $1.05\,M_{\mathrm{Jup}}$ (ATMO 2020 and CBPD) or $1.05$–$3.14\,M_{\mathrm{Jup}}$ (Sonora Bobcat, Sonora Diamondback, and SM08). By comparison, the standard public BT-Settl evolutionary tables of \citet{btsettle2014} and the \citet{baraffe2015} tracks have intervals of $\simeq10.5\,M_{\mathrm{Jup}}$ across this range. In Table~\ref{tab:evolutionary_models}, we list a summary of the available evolutionary models and some of their core characteristics.

\begingroup
\setlength{\tabcolsep}{2pt}
\renewcommand{\arraystretch}{1.15}
\newcommand{\evolcell}[2]{\parbox[t]{#1\columnwidth}{\raggedright #2\strut}}
\begin{deluxetable}{@{}lllll@{}}[htbp]
\tabletypesize{\scriptsize}
\tablewidth{\columnwidth}
\tablecaption{Atmospheric and evolutionary models\label{tab:evolutionary_models}}
\tablehead{
\colhead{Model} & \colhead{Spectra} & \colhead{$[\mathrm{M/H}]$} & \colhead{Cloud treatment} & \colhead{Modelling notes}
}
\startdata
\evolcell{0.20}{ATMO 2020} & \evolcell{0.11}{CEQ, NEQ} & \evolcell{0.12}{$0$} & \evolcell{0.21}{Cloudless} & \evolcell{0.28}{Different CEQ/NEQ spectra; shared cooling tracks.} \\[3pt]
\evolcell{0.20}{CBPD} & \evolcell{0.11}{CEQ, NEQ} & \evolcell{0.12}{$0$} & \evolcell{0.21}{Cloudless} & \evolcell{0.28}{ATMO spectra; updated H/He EOS changes cooling.} \\[3pt]
\evolcell{0.20}{Sonora Bobcat} & \evolcell{0.11}{CEQ} & \evolcell{0.12}{$0,\,\pm0.5$} & \evolcell{0.21}{Cloudless} & \evolcell{0.28}{Opacity differences; metallicity changes spectra and cooling.} \\[3pt]
\evolcell{0.20}{Sonora Diamondback} & \evolcell{0.11}{CEQ} & \evolcell{0.12}{$0,\,\pm0.5$} & \evolcell{0.21}{Hybrid or hybrid-grav; $f_{\mathrm{sed}}=2$} & \evolcell{0.28}{Updated opacities and EOS; gravity-dependent clearing available.} \\[3pt]
\evolcell{0.20}{SM08} & \evolcell{0.11}{CEQ} & \evolcell{0.12}{$0$} & \evolcell{0.21}{Cloudy or hybrid; $f_{\mathrm{sed}}=2$} & \evolcell{0.28}{Fixed cloudy boundary method, or hybrid cloud clearing over $1400$--$1200\,\mathrm{K}$.}
\enddata
\tablecomments{CEQ and NEQ denote chemical-equilibrium and disequilibrium spectra. In ATMO and CBPD, weak/strong NEQ mixing changes the spectra but uses the corresponding CEQ cooling tracks. ``Hybrid'' denotes cooling that transitions from cloudy to cloudless atmospheric spectra as the brown dwarf cools through the L/T transition.}
\end{deluxetable}
\endgroup

In reference to Table~\ref{tab:evolutionary_models}, the ATMO 2020 models with weak and strong vertical mixing change the median temperatures and ages by at most $8\,\mathrm{K}$ and $31\,\mathrm{Myr}$ relative to chemical equilibrium, with nearly unchanged uncertainties. We therefore omit these results. Our bandpass-integrated eclipse depths have negligible sensitivity to these spectral changes, which could be explored with higher-resolution spectroscopy. We also omit the solar-metallicity cloudless Sonora Diamondback results, which agree with ATMO 2020 and Sonora Bobcat within the inferred uncertainties, yielding median temperature and age differences of $\sim 10\,\mathrm{K}$ and $20\,\mathrm{Myr}$. We also do not evaluate the SM08 cloudless, chemical-equilibrium models at $\mathrm{[M/H]}=-0.3,0.0,+0.3$ for our comparison.

\subsection{Intrinsic Temperature Calculation}\label{sec:internal_temp}

\begin{figure*}[t!]
    \centering
    \includegraphics[width=\textwidth]{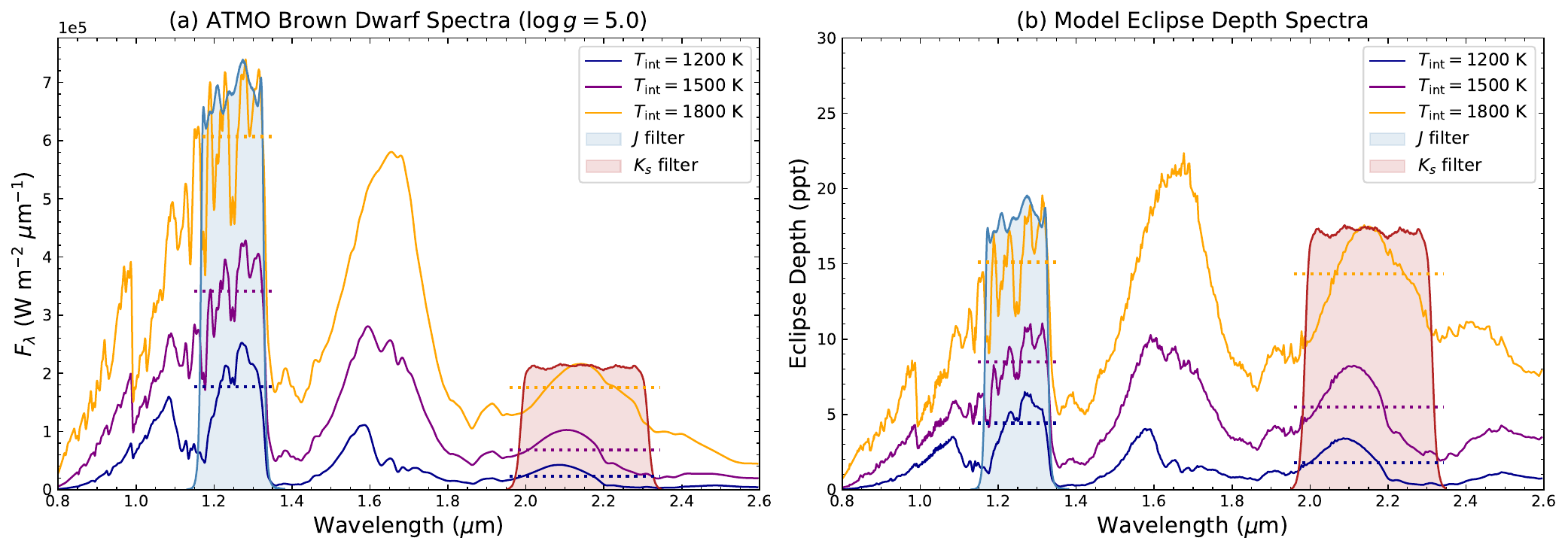}
    \caption{(a) ATMO 2020 model spectra at $\log g = 5.0$ and three intrinsic temperatures $T_{\mathrm{int}} = 1200$, $1500$, and $1800$ K (dark blue, purple, and orange, respectively), smoothed to a spectral resolution of $R = 200$. The $J$ and $K_s$ filter transmission curves (blue and red shaded regions) are overlaid, scaled to the peak passband flux of the $T_{\mathrm{int}} = 1800$ K spectrum. Dotted horizontal lines indicate the band-averaged flux within each filter. (b) Model secondary-eclipse depth spectra in parts per thousand (ppt) using a PHOENIX stellar atmosphere model \citep{husser2013} with parameters $T_{\mathrm{eff},\star} = 3200$ K and $\log g_\star = 5.0$ (cgs). These sample eclipse depth spectra use $R_{\mathrm{BD}}/R_\star = 0.30$. Dotted horizontal lines indicate the band-averaged eclipse depths.}
    \label{fig:atmo_eclipse_depths}
\end{figure*}

To estimate the ages of the two systems, we first inferred the intrinsic effective temperatures of the brown dwarf companions, allowing us to place them on brown dwarf cooling tracks. We constrained these temperatures by comparing the secondary-eclipse depths measured from our joint \texttt{juliet} fits with synthetic eclipse depths generated using the coupled evolutionary and atmospheric models. For a given model intrinsic temperature, we calculated the predicted secondary-eclipse depth in the instrumental bandpass as:
\begin{equation}
\label{equation_eclipse_depth}
\delta_{\mathrm{ecl}}
=
\left(\frac{R_{\mathrm{BD}}}{R_{\star}}\right)^{2}
\frac{
\int F_{\lambda,\mathrm{BD}}\,
\mathcal{T}(\lambda)\lambda\,d\lambda
}{
\int F_{\lambda,\star}\,
\mathcal{T}(\lambda)\lambda\,d\lambda
},
\end{equation}
where $F_{\lambda,\mathrm{BD}}$ and $F_{\lambda,\star}$ are the emergent surface flux densities of the brown dwarf and host star, respectively, $R_{\mathrm{BD}}$ and $R_{\star}$ are their radii, and $\mathcal{T}(\lambda)$ is the filter transmission function of WIRC in the observed passband. The additional factor of $\lambda$ converts the model spectral energy flux densities to photon counts as $E_{photon} = hc/\lambda$.

Figure~\ref{fig:atmo_eclipse_depths} illustrates ATMO 2020 spectra and their corresponding eclipse depths for $T_{\mathrm{int}}=1200$, $1500$, and $1800\,\mathrm{K}$ at $\log g=5.0$. Panel (a) shows the emergent surface flux densities: hotter models produce greater near-infrared flux, although molecular absorption lines can vary due to atmospheric temperature and chemistry. For example, the absorption feature near $1.7\,\mu\mathrm{m}$ weakens as $T_{\mathrm{int}}$ increases. Panel (b) shows the wavelength-dependent eclipse depth, $\delta_\lambda=(R_{\mathrm{BD}}/R_\star)^2F_{\lambda,\mathrm{BD}}/F_{\lambda,\star}$, calculated after interpolating the stellar spectrum onto the brown dwarf wavelength grid. Although the brown dwarf emits less flux in $K_s$ than in $J$, the hotter star’s declining flux toward longer near-infrared wavelengths offsets this difference, producing comparable eclipse depths in the $J$, $H$, and $K_s$ bands. Dotted lines indicate the filter-weighted averages. For the adopted stellar spectrum and radius ratio, the $1500\,\mathrm{K}$ model predicts eclipse depths of approximately $8.5$ and $6.0\,\mathrm{ppt}$ in $J$ and $K_s$, respectively, illustrating how these measurements can constrain $T_{\mathrm{int}}$.

Because the model spectra are tabulated in $T_{\mathrm{int}}$ and $\log g$, we use a two-step interpolation. We first evaluate the photon-weighted passband integral $I(T_{\mathrm{int}},\log g)=\int F_{\lambda,\mathrm{BD}}\mathcal{T}(\lambda)\lambda\,d\lambda$ at each grid temperature and at two $\log g$ values that bound the median gravity. These bounds are $\log g=5.0$ and $5.5$, except for SM08, where we use grid points $5.00$ and $5.25$ for AD 3116B and $5.25$ and $5.48$ for LP 261-75C. At both the lower and upper gravity bounds, we interpolate the integral values in $T_{\mathrm{int}}$ using cubic splines, yielding two continuous curves of flux integral versus temperature. We then interpolate these two curves in $\log g$ to each system's median gravity. This defines a continuous function $I(T_{\mathrm{int}})$ for our eclipse-depth calculation. For the ATMO 2020 equilibrium models, perturbing $\log g$ by its $1\sigma$ uncertainties changes the inferred temperature by less than $3\,\mathrm{K}$ for both systems. These shifts are much smaller than the inferred temperature uncertainties, so we do not propagate the uncertainty in $\log g$.

We evaluate the stellar passband integral $F_\star=\int F_{\lambda,\star}\mathcal{T}(\lambda)\lambda\,d\lambda$ using the IRTF spectrum of LP 261-75 in $K_s$ and the SPHEREx spectrum of AD 3116 in $J$, both expressed as stellar surface flux densities (Figure~\ref{fig:stellar_surface_spectra}). We linearly interpolated the filter transmission onto each spectrum's wavelength grid including the spectral measurement and geometric scaling uncertainties described in Section~\ref{sec:3pt7}. We find photon-weighted stellar passband integrals of $(8.55^{+0.53}_{-0.49})\times10^5\,\mathrm{W\,m^{-2}}$ from IRTF/SpeX for LP 261-75A in $K_s$ and $(8.10^{+0.75}_{-0.66})\times10^5\,\mathrm{W\,m^{-2}}$ from SPHEREx for AD 3116A in $J$. For LP 261-75A, SPHEREx gives $(8.43^{+0.53}_{-0.48})\times10^5\,\mathrm{W\,m^{-2}}$, only 1.4\% lower, showing remarkable agreement despite the large difference in spectral resolution. We then defined a Gaussian likelihood for the measured eclipse depths and sampled the joint posterior in $R_{\mathrm{BD}}/R_\star$, $F_\star$, and $T_{\mathrm{int}}$ using the MCMC sampler \texttt{emcee} \citep{foreman2013}. We imposed independent Gaussian priors on the radius ratio and stellar passband integral, based on the global orbital fits and observed stellar spectra. We adopted a uniform temperature prior over $750$--$3000\,\mathrm{K}$. We used 32 walkers for 10,000 steps, discarding the first 3,000 steps as burn-in. This yielded posterior distributions and covariances for the three parameters. We subsequently used the temperature posterior for age estimation in Section~\ref{sec:4pt5}.

\subsection{Assumptions on Stellar Irradiation}\label{sec:tint_and_irradiation}

Our calculation thus far has assumed that stellar irradiation contributes negligibly to heating the brown dwarf's atmosphere. In the presence of stellar irradiation, however, the emitted thermal flux includes both the intrinsic flux and absorbed stellar radiation \citep{fortney2020, mukherjee2026},
\begin{equation}
T_{\mathrm{eff}}^4 = T_{\mathrm{int}}^4 + T_{\mathrm{eq}}^4,
\end{equation}
where $T_{\mathrm{int}}$ describes the intrinsic flux from the interior, $T_{\mathrm{eq}}$ describes the contribution from stellar heating, and $T_{\mathrm{eff}}$ corresponds to the total emitted thermal flux. Thus, the temperature inferred from our eclipse depths can include a contribution from stellar irradiation. We denote this fitted temperature as $T_{\mathrm{fit}}$ to distinguish it from the intrinsic temperature used in the cooling models.

Using the stellar and orbital parameters in Tables 1--4, we compute $T_{\mathrm{eq}}=575\pm29$ and $571\pm29\,\mathrm{K}$ for LP 261-75C and AD 3116B, respectively, assuming zero Bond albedo (complete absorption) and reradiation over the full surface via efficient heat transport from the dayside to the nightside. Assuming reradiation over only the dayside instead gives $T_{\mathrm{eq}}=684\pm34$ and $679\pm35\,\mathrm{K}$. For ATMO 2020 fitted temperatures of $1591\pm24$ (LP~261-75C) and $1475\pm60\,\mathrm{K}$ (AD~3116B), these imply irradiation contributions of $(T_{\mathrm{eq}}/T_{\mathrm{fit}})^4 \simeq 1.7\%$ and $2.3\%$ for full-surface redistribution, increasing to $3.4\%$ and $4.6\%$ for dayside reradiation.

Because stellar irradiation contributes only a small fraction of the emitted flux, we expect its effect on the atmospheric spectra and cooling evolution to be small. This is consistent with the irradiated evolutionary calculations of \cite{mukherjee2026}, which predict nearly identical radii for both targets with and without irradiation. To account for the additional thermal emission, we compute three cases: no irradiation correction, full-surface redistribution ($T_{\mathrm{eq}}\approx570\,\mathrm{K}$), and dayside reradiation ($T_{\mathrm{eq}}\approx680\,\mathrm{K}$). Approximating $T_{\mathrm{fit}}\simeq T_{\mathrm{eff}}$, we estimate the intrinsic temperature as
\begin{equation}
T_{\mathrm{int}} \simeq
\left(T_{\mathrm{fit}}^4-T_{\mathrm{eq}}^4\right)^{1/4},
\end{equation}
and use these corrected temperatures, including the uncertainty in $T_{\mathrm{eq}}$, to infer ages from the evolutionary models.

Another consideration for the inferred temperature is reflected starlight, which directly adds to the thermal eclipse depth in Equation~\ref{equation_eclipse_depth}. Near full phase, this contribution is $\delta_{\mathrm{refl}}\simeq A_{\mathrm{g,band}}(R_{\mathrm{BD}}/a)^2$. For the eccentric orbit of AD~3116B, the separation at secondary eclipse is $r_{\mathrm{sec}}\simeq1.009a$, so using the semimajor axis is sufficient for estimating the reflected-light contribution. Here, $A_{\mathrm{g,band}}$ is the geometric albedo in the observed passband, describing the reflected brightness at full phase relative to a perfectly reflecting, diffusely scattering disk of the same projected area \citep{winn2010}. The Bond albedo instead measures the fraction of incident stellar energy reflected over all wavelengths and directions. The two are related through the phase integral, which accounts for the angular distribution of reflected light, and an average over wavelength weighted by the incident stellar flux \citep{burrowsorton2010,schwartz2015}.

Although we cannot determine the geometric albedo from our data, we can bound its systematic effect. Optical eclipse and phase-curve measurements find geometric albedos of order $0.1$ for many hot Jupiters \citep{schwartz2015,mallonn2019}. These measurements do not necessarily correspond to reflection in our $J$ and $K_s$ passbands, so we consider an illustrative hyperbolic value of $A_{\mathrm{g,band}}=0.5$. Using our median radii and semimajor axes, the reflected-light contributions are $\delta_{\mathrm{refl},K_s}\simeq390\,A_{\mathrm{g},K_s}\,\mathrm{ppm}$ for LP 261-75C and $\delta_{\mathrm{refl},J}\simeq501\,A_{\mathrm{g},J}\,\mathrm{ppm}$ for AD 3116B. At $A_{\mathrm{g,band}}=0.5$, these corrections become  $195$ and $250\,\mathrm{ppm}$, corresponding to $0.65\sigma$ of the LP 261-75C eclipse-depth uncertainty and $0.44\sigma$ of the weighted AD~3116B eclipse-depth uncertainty.

Subtracting these contributions and recomputing the ATMO 2020 fits reduces the median fitted temperature from $1591$ to $1581\,\mathrm{K}$ for LP 261-75C and from $1475$ to $1462\,\mathrm{K}$ for AD 3116B, with nearly unchanged posterior uncertainties of $\simeq$ $24$ and $60\,\mathrm{K}$. When computing the stellar radiation impact via equilibrium temperatures, we assumed zero Bond albedo. This maximizes the absorbed stellar energy and corresponds to $A_g \approx 0$. Therefore, increasing the fraction of reflected light would reduce the energy available for atmospheric heating and lower $T_{\mathrm{eq}}$, although the magnitude of this reduction cannot be determined from a geometric albedo in one passband alone \citep{schwartz2015}. Given the small temperature shifts found above, we retain the complete-absorption assumption for our stellar irradiation corrections and neglect reflected light in our temperature inference. This exercise on the geometric albedo and its competing contribution with $T_{eq}$ gives us confidence that our uncertainties are appropriate, and that our results do not depend sensitively on the M-dwarf's stellar radiation. 

\subsection{Contribution of Tidal Heating}

Eccentricity-damping tidal dissipation can contribute to elevated internal temperatures in close-in planets and brown dwarfs. The contribution is negligible for LP~261-75C, which already has a circular orbit, but is likely non-zero for AD~3116B, which we measure to have an orbital eccentricity of $e \approx 0.14$. To estimate the contribution of tidal heating to the overall intrinsic temperature, we use the following expression for heating from equilibrium tides in a synchronously rotating object from \citet{jackson2008tidalheatingB}:
\begin{equation}
    \dot{E} = \frac{63}{4} \frac{(G M_\star)^{3/2} M_\star R_2^5}{Q^\prime} a^{-15/2} e^2,
\end{equation}
where $Q^\prime$ is the tidal quality factor of the brown dwarf. Converting this heating rate to an intrinsic temperature and rearranging gives us
\begin{equation} 
\label{eq:tidal_temp}
\begin{split}
      T_{\rm int,tides} \approx 1000 \, {\rm K} \left(\frac{Q^\prime}{10^5}\right)^{-1/4} \left(\frac{M_\star}{0.27 \, M_\odot}\right)^{5/8} \\ 
      \times \left(\frac{R_2}{R_{\rm Jup}}\right)^{3/4} \left(\frac{a}{0.02 \, {\rm au}}\right)^{-15/8} \left(\frac{e}{0.14}\right)^{1/2}  .
\end{split}
\end{equation}
Several studies have estimated the value of $Q^\prime$ for giant planets and have generally found the value to range between $10^5 - 10^7$ \citep[e.g.,][]{jackson2008tidalheatingA, lainey2009jupitertides, bonomo2017tidalQ, giacalone2017tidalQ}. However, \citet{ferreira2026tides} recently showed that $Q^\prime$ for brown dwarfs is likely higher than it is for giant planets, possibly even reaching $10^8$. Conservatively assuming $Q^\prime = 10^6$, we obtain $T_{\rm int,tides} \approx 560$~K, which is comparable to the heating from stellar irradiation discussed above. If we instead assume $Q^\prime = 10^8$, we get $T_{\rm int,tides} \approx 180$~K. Thus, tidal heating likely accounts for $\lesssim 2\%$ of the total intrinsic temperature of the object, making it negligible for the purposes of our analysis.

\subsection{Age Estimation}\label{sec:4pt5}

The cooling models natively tabulate $(M_{\mathrm{BD}},\mathrm{age})\rightarrow T_{\mathrm{int}}$ along fixed-mass tracks. Temperature decreases with age, although deuterium burning can temporarily slow cooling at young ages. For ATMO 2020, each track contains 40 logarithmically spaced ages from $1\,\mathrm{Myr}$ to $10\,\mathrm{Gyr}$, with successive ages separated by a factor of $1.26$. To infer the ages for our systems, we use a root solver to invert this mapping — using the mass constraints from our global orbital fits and the intrinsic-temperature constraints from our eclipse-depth analysis.

\begin{figure}[ht!]
    \centering
    \includegraphics[width=\columnwidth]{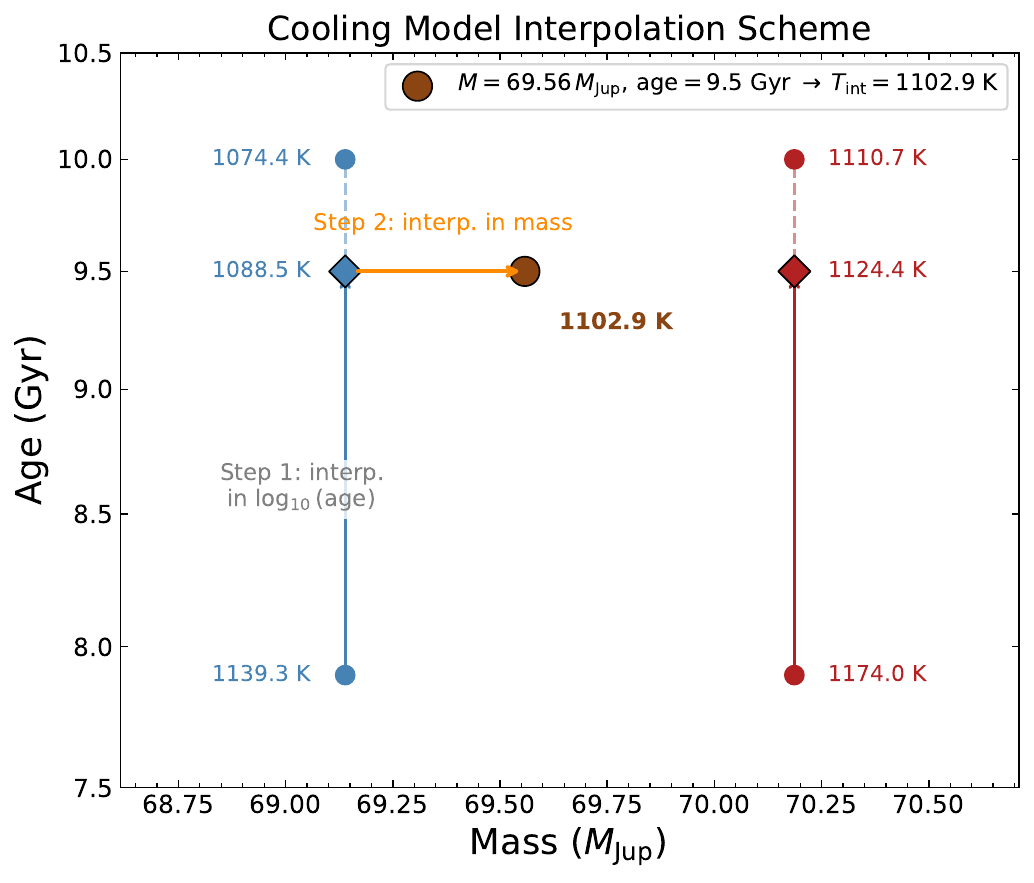}
    \caption{Two-step interpolation of $(M_{\mathrm{BD}},\mathrm{age})\rightarrow T_{\mathrm{int}}$, illustrated at $M_{\mathrm{BD}}=69.56\,M_{\mathrm{Jup}}$ and $\mathrm{age}=9.5\,\mathrm{Gyr}$. Blue and red circles mark native grid points along the bracketing $0.066$ and $0.067\,M_\odot$ tracks. Step 1 interpolates temperature linearly in $\log_{10}(\mathrm{age})$ along each track, yielding the diamonds at the target age. Step 2 interpolates linearly in mass between these values, yielding $T_{\mathrm{int}}=1102.9\,\mathrm{K}$ (brown circle). This calculation is evaluated repeatedly by the root solver to infer age from $(M_{\mathrm{BD}},T_{\mathrm{int}})$.}
    \label{fig:cooling_interpolation}
\end{figure}

Following the models' native mapping $(M_{\mathrm{BD}},\mathrm{age})\rightarrow T_{\mathrm{int}}$, we created a forward model that continuously predicts $T_{\rm{int}}$ for any value of $M_{\rm{BD}}$ and $\rm{age}$ by interpolating within the grid bounds. Figure~\ref{fig:cooling_interpolation} illustrates this calculation, which evaluates $T_{\mathrm{int}}$ at a specified $(M_{\mathrm{BD}},\mathrm{age})$. For each trial pair, we identify the bracketing mass tracks, $M_{\mathrm{l}}\leq M_{\mathrm{BD}}\leq M_{\mathrm{u}}$. Along each track, we interpolate temperature linearly in $\log_{10}(\mathrm{age})$ between the neighboring grid ages, $\mathrm{age}_{\mathrm{l}}$ and $\mathrm{age}_{\mathrm{u}}$, that bracket the trial age. This gives the temperatures $T_l(\mathrm{M_l},\mathrm{age})$ and $T_u(\mathrm{M_u},\mathrm{age})$ (Step 1). We then interpolate temperature linearly in mass between these two values to obtain $T_{\mathrm{int}}$ at the desired $M_{\mathrm{BD}}$ (Step 2), completing the continuous mapping.

We subsequently invert this forward model to infer age from our data of $(M_{\mathrm{BD}},T_{\mathrm{int}})$. For each sampled pair $(M_{\mathrm{BD},i},T_{\mathrm{int},i})$, a root solver holds the sampled mass fixed and varies the trial age until $T_{\mathrm{int,model}}(M_{\mathrm{BD},i},\mathrm{age}_i)=T_{\mathrm{int},i}$. We repeat this calculation for $10{,}000$ independent draws from Gaussian realizations of the mass and temperature distributions, yielding our system ages. A sample cooling track is displayed in Figure~\ref{fig:lp261_75c_schematic} of Section~\ref{sec:5pt2}.

\section{Results}\label{sec:results}
\subsection{Inferred Temperatures and Ages}

\begin{deluxetable*}{llccc|ccc}
\tabletypesize{\footnotesize}
\renewcommand{\arraystretch}{1.2}
\tablewidth{0pt}
\tablecaption{Intrinsic temperatures and evolutionary ages under different irradiation assumptions \label{tab:evolutionary_comparison}}
\tablehead{
\colhead{Evolutionary model} & \colhead{Quantity} & \multicolumn{3}{c|}{LP~261-75C} & \multicolumn{3}{c}{AD~3116B} \\
\colhead{} & \colhead{$T_{\mathrm{eq}}$ (K)} & \colhead{$0$} & \colhead{$575 \pm 29$} & \multicolumn{1}{c|}{$684 \pm 34$} & \colhead{$0$} & \colhead{$571 \pm 29$} & \colhead{$679 \pm 35$}
}
\startdata
ATMO 2020 / CBPD & $T_{\mathrm{int}}$ (K) & $1591 \pm 24$ & $1584 \pm 24$ & $1577 \pm 24$ & $1475 \pm 60$ & $1465 \pm 62$ & $1457 \pm 63$ \\
ATMO 2020 (EOS 2019) & Age (Myr) & ${1690}^{+400}_{-280}$ & ${1720}^{+410}_{-290}$ & ${1750}^{+420}_{-300}$ & ${1140}^{+290}_{-240}$ & ${1160}^{+300}_{-250}$ & ${1180}^{+310}_{-250}$ \\
CBPD (EOS 2021) & Age (Myr) & ${1490}^{+380}_{-220}$ & ${1510}^{+330}_{-220}$ & ${1530}^{+330}_{-230}$ & ${1010}^{+250}_{-200}$ & ${1030}^{+250}_{-210}$ & ${1060}^{+260}_{-210}$ \\[3pt]
\hline
\noalign{\vskip 3pt}
\multirow{2}{*}{Sonora Bobcat, $[\mathrm{M/H}]=+0.5$} & $T_{\mathrm{int}}$ (K) & $1513 \pm 20$ & $1505 \pm 21$ & $1496 \pm 21$ & $1487 \pm 57$ & $1478 \pm 59$ & $1469 \pm 60$ \\
 & Age (Myr) & ${2810}^{+1410}_{-710}$ & ${2880}^{+1520}_{-710}$ & ${2940}^{+1610}_{-720}$ & ${1270}^{+340}_{-310}$ & ${1300}^{+360}_{-320}$ & ${1330}^{+370}_{-320}$ \\[3pt]
\multirow{2}{*}{Sonora Bobcat, $[\mathrm{M/H}]=0.0$} & $T_{\mathrm{int}}$ (K) & $1582 \pm 22$ & $1575 \pm 23$ & $1567 \pm 23$ & $1454 \pm 56$ & $1441 \pm 58$ & $1435 \pm 59$ \\
 & Age (Myr) & ${1780}^{+510}_{-320}$ & ${1810}^{+530}_{-330}$ & ${1840}^{+550}_{-340}$ & ${1210}^{+290}_{-280}$ & ${1240}^{+300}_{-290}$ & ${1270}^{+310}_{-300}$ \\[3pt]
\multirow{2}{*}{Sonora Bobcat, $[\mathrm{M/H}]=-0.5$} & $T_{\mathrm{int}}$ (K) & $1671 \pm 25$ & $1664 \pm 26$ & $1659 \pm 26$ & $1423 \pm 50$ & $1413 \pm 52$ & $1403 \pm 53$ \\
 & Age (Myr) & ${1310}^{+270}_{-220}$ & ${1320}^{+280}_{-220}$ & ${1340}^{+280}_{-220}$ & ${1200}^{+270}_{-270}$ & ${1230}^{+280}_{-280}$ & ${1260}^{+290}_{-290}$ \\[3pt]
\hline
\noalign{\vskip 3pt}
Sonora Diamondback, $[\mathrm{M/H}]=+0.5$ & $T_{\mathrm{int}}$ (K) & $1430 \pm 25$ & $1420 \pm 26$ & $1410 \pm 27$ & $1673 \pm 36$ & $1667 \pm 37$ & $1661 \pm 37$ \\
\quad Hybrid & Age (Myr) & ${2060}^{+790}_{-530}$ & ${2120}^{+810}_{-560}$ & ${2170}^{+890}_{-580}$ & ${500}^{+120}_{-100}$ & ${500}^{+120}_{-100}$ & ${510}^{+120}_{-100}$ \\
\quad Hybrid-grav & Age (Myr) & ${1560}^{+540}_{-300}$ & ${1600}^{+640}_{-300}$ & ${1720}^{+770}_{-380}$ & ${420}^{+100}_{-70}$ & ${420}^{+100}_{-80}$ & ${430}^{+100}_{-80}$ \\[3pt]
Sonora Diamondback, $[\mathrm{M/H}]=0.0$ & $T_{\mathrm{int}}$ (K) & $1434 \pm 29$ & $1424 \pm 30$ & $1414 \pm 31$ & $1663 \pm 37$ & $1657 \pm 38$ & $1651 \pm 38$ \\
\quad Hybrid & Age (Myr) & ${1840}^{+720}_{-370}$ & ${1880}^{+770}_{-390}$ & ${1940}^{+800}_{-410}$ & ${530}^{+110}_{-100}$ & ${540}^{+120}_{-100}$ & ${540}^{+120}_{-100}$ \\
\quad Hybrid-grav & Age (Myr) & ${1790}^{+640}_{-390}$ & ${1850}^{+740}_{-410}$ & ${1930}^{+870}_{-430}$ & ${510}^{+110}_{-90}$ & ${520}^{+110}_{-100}$ & ${520}^{+110}_{-100}$ \\[3pt]
Sonora Diamondback, $[\mathrm{M/H}]=-0.5$ & $T_{\mathrm{int}}$ (K) & $1451 \pm 30$ & $1441 \pm 31$ & $1432 \pm 32$ & $1640 \pm 36$ & $1633 \pm 37$ & $1627 \pm 37$ \\
\quad Hybrid & Age (Myr) & ${1580}^{+450}_{-270}$ & ${1610}^{+470}_{-280}$ & ${1660}^{+480}_{-290}$ & ${570}^{+120}_{-100}$ & ${570}^{+120}_{-100}$ & ${580}^{+120}_{-100}$ \\
\quad Hybrid-grav & Age (Myr) & ${1600}^{+460}_{-280}$ & ${1660}^{+480}_{-310}$ & ${1720}^{+510}_{-350}$ & ${570}^{+120}_{-100}$ & ${570}^{+120}_{-100}$ & ${580}^{+120}_{-100}$ \\[3pt]
\hline
\noalign{\vskip 3pt}
\multirow{2}{*}{Saumon \& Marley (2008), hybrid} & $T_{\mathrm{int}}$ (K) & $1427 \pm 24$ & $1416 \pm 25$ & $1407 \pm 26$ & $1687 \pm 27$ & $1681 \pm 28$ & $1675 \pm 28$ \\
 & Age (Myr) & ${1940}^{+840}_{-380}$ & ${1990}^{+900}_{-360}$ & ${2150}^{+860}_{-440}$ & ${500}^{+90}_{-110}$ & ${500}^{+90}_{-110}$ & ${510}^{+90}_{-110}$ \\
\enddata
\tablecomments{The equilibrium temperatures used for each system are listed in the column headers. For each system, from left to right, the three columns correspond to no irradiation ($T_{\mathrm{eq}}=0$), re-radiation over the full surface, and dayside-only re-radiation. Entries give median intrinsic temperatures and ages, with age uncertainties spanning the 16th–84th percentiles of the posterior distributions. Ages are rounded to the nearest 10 Myr. ATMO 2020 and CBPD share temperature inputs. At each Diamondback metallicity, the temperature value applies to both hybrid prescriptions, with their separate ages listed immediately below. The SM08 model uses hybrid cooling with a cloudy spectral boundary condition.}
\end{deluxetable*}

Table~\ref{tab:evolutionary_comparison} summarizes our inferred intrinsic temperatures and evolutionary ages. We tabulate results for ATMO 2020, CBPD, Sonora Bobcat, Sonora Diamondback, and SM08, including the metallicity and cloud-clearing variants described in Section~\ref{sec:evolutionary_models_evaluation}. For each model, we report results assuming no stellar irradiation, redistribution over the full surface, and dayside-only reradiation.

The two systems exhibit different sensitivities to model choice, which can arise from a combination of their differing masses and observed passbands. Across the tabulated cases, LP~261-75C has inferred intrinsic temperatures of $1400$--$1700$~K and median ages of $1.3$--$2.9$~Gyr. All models predict ages significantly older than the $130$--$200$~Myr estimate based on proposed membership in the AB Doradus moving group \citep{irwin2018lp26175}. For AD~3116B, the models yield intrinsic temperatures of approximately $1400$--$1700$~K and median ages of $0.4$--$1.3$~Gyr. Mass uncertainty dominates the age uncertainty for LP~261-75C, which we quantify by fixing either the mass or $T_{\mathrm{int}}$ at its median while propagating the other quantity's uncertainty through a Monte Carlo procedure. For LP 261-75C, the ATMO 2020 model yields an age standard deviation of $400$~Myr from the mass uncertainty, compared to $94$~Myr from the temperature uncertainty. For AD~3116B, the contributions are more comparable ($221$ and $182$~Myr, respectively) due to the less precise eclipse-depth measurement.

Stellar irradiation produces modest age shifts relative to the inferred uncertainties. Adopting the dayside-only correction lowers the median intrinsic temperatures by $12$--$20$~K compared with the no-irradiation case. The corresponding median ages increase by at most $11\%$ for LP~261-75C and $5\%$ for AD~3116B. Full-surface redistribution produces noticeably smaller shifts. As the fractional contribution of irradiation to the total flux scales as $(T_{\mathrm{eq}}/T_{\mathrm{fit}})^4$, the impact of $\Delta T$ is smaller at higher fitted temperatures. Unless otherwise stated, the model comparisons below refer to the case of no stellar irradiation.

The cloudy versus cloudless model families shift the inferred temperatures in opposite directions for the two systems. The solar metallicity Diamondback model gives a temperature for LP~261-75C that is $\sim 150$~K lower than the non-cloudy Bobcat model, although the two models agree on an age of about $1800$~Myr. For AD~3116B, the cloudy Diamondback model gives a temperature that is $\sim 200$~K higher than the cloudless Bobcat model and a substantially younger median age by $700$~Myr.

Regarding metallicity, as $[\mathrm{M/H}]$ decreases from $+0.5$ to $-0.5$, the Bobcat temperature rises from $1513$ to $1671$~K for LP~261-75C but falls from $1487$ to $1423$~K for AD~3116B. Metallicity most strongly affects the LP~261-75C ages inferred with Bobcat: increasing $[\mathrm{M/H}]$ from $-0.5$ to $+0.5$ raises the median age from $1310$ to $2810$~Myr, a factor of $2.1$. The Diamondback hybrid ages for LP~261-75C also increase with metallicity, although the dependence is weaker. Assuming that LP~261-75C shares the composition of its host star, our abundance measurement of $\mathrm{[Fe/H]}\gtrsim+0.5$ for LP~261-75A (Section~\ref{sec:irtf}) favors an older age within the Bobcat and Diamondback model grids. AD~3116B is instead minimally sensitive to metallicity, with Bobcat and Diamondback hybrid median ages respectively spanning $1200$--$1270$ and $500$--$570$~Myr.

The updated EOS reduces the inferred ages. Relative to ATMO 2020, CBPD lowers the median age from $1690$ to $1490$~Myr for LP~261-75C and from $1140$ to $1010$~Myr for AD~3116B, reductions of $\approx12\%$ and $\approx11\%$. These shifts remain within the age uncertainties. The solar-metallicity Bobcat results also agree broadly with both cloudless model sets, yielding temperatures within $20$~K of the ATMO/CBPD values for both systems.

The Diamondback hybrid and hybrid-gravity prescriptions yield similar ages except at super-solar metallicity. This agreement is expected, as both prescriptions implement cloud clearing at temperatures below those inferred for our systems. At $[\mathrm{M/H}]=+0.5$, replacing hybrid cooling with hybrid-gravity cooling reduces the median age from $2060$ to $1560$~Myr for LP~261-75C and from $500$ to $420$~Myr for AD~3116B.

\begin{figure*}[t!]
    \centering
    \includegraphics[width=\textwidth]{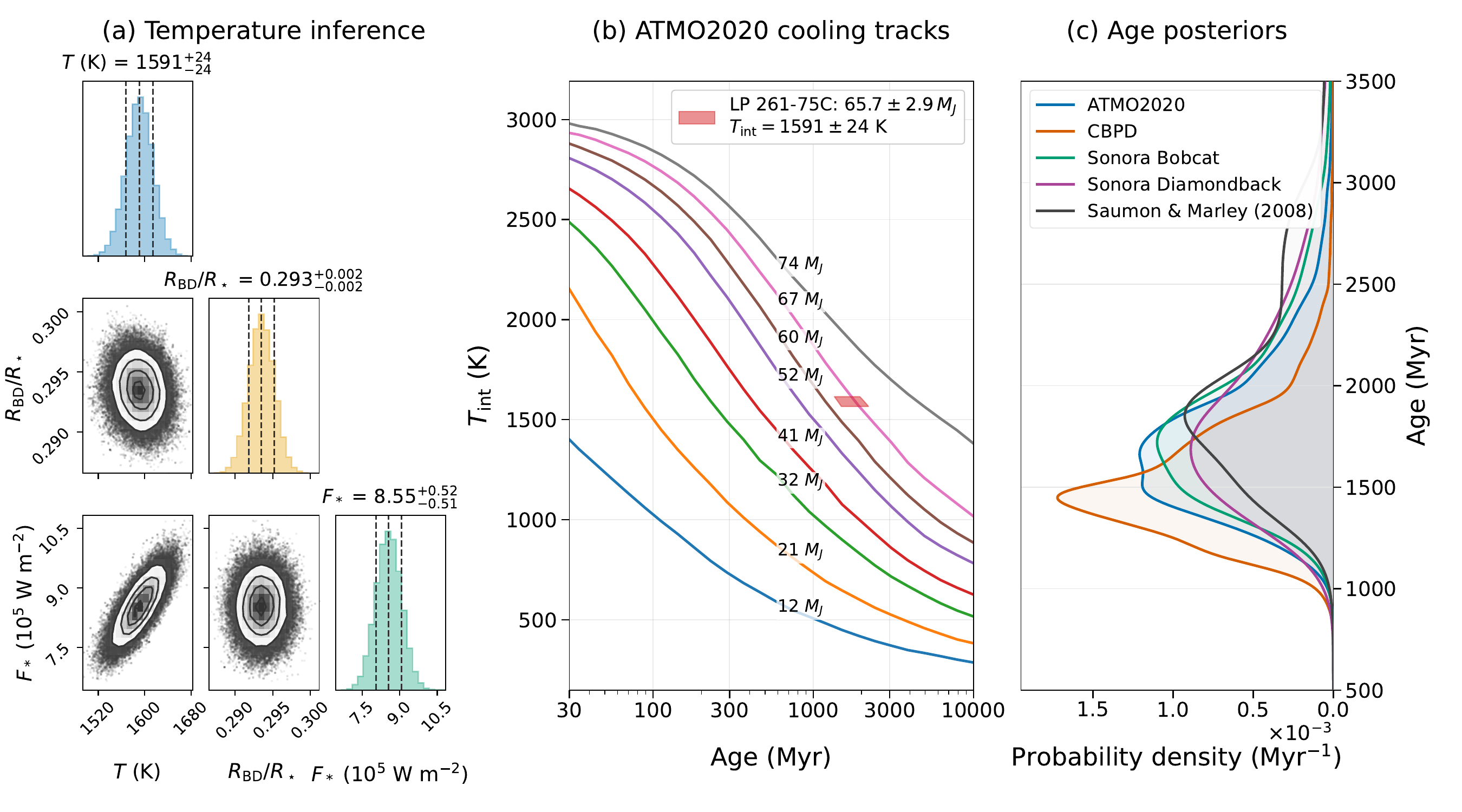}
    \caption{Temperature and age constraints for LP 261-75C. (a) Marginalized and joint posterior distributions from our MCMC of the intrinsic temperature $T_{\mathrm{int}}$, radius ratio $R_{\mathrm{BD}}/R_\star$, and integrated stellar passband flux $F_\star$. (b) ATMO2020 chemical-equilibrium (CEQ) cooling tracks. The red region highlights the $1\sigma$ uncertainties on the temperature inferred from the posterior distribution and the system mass, yielding an age of $1690^{+400}_{-280}$ Myr. (c) Age posterior distributions for solar-metallicity evolutionary models, with Sonora Diamondback evaluated with hybrid cooling.}
    \label{fig:lp261_75c_schematic}
\end{figure*}

The Diamondback ages lie closest to the independently estimated Praesepe age range of $600$--$800$~Myr (see Section~\ref{ad3116_subsection}). The solar-metallicity hybrid result for AD~3116B, $530^{+110}_{-100}$~Myr, has a 68\% credible interval that overlaps the lower end of this $600$--$800$~Myr range. The subsolar-metallicity prescriptions yield the closest median age, $570^{+120}_{-100}$~Myr. Despite giving closer median ages, we note that Praesepe is metal rich with $[\mathrm{Fe/H}]\simeq+0.1$--$+0.3$ (Section~\ref{ad3116_subsection}), making the subsolar models less representative. Among the cloudless models, CBPD gives the closest median age, $1010^{+250}_{-200}$~Myr, and its lower 68\% bound lies at the top of the adopted cluster-age range.

\subsection{Cooling Tracks and Eclipse-Depth Predictions}\label{sec:5pt2}
In Figure~\ref{fig:lp261_75c_schematic}, we present a schematic of our framework, illustrating our MCMC, cooling model evaluation, and age inference. The corner plot in panel (a) illustrates how uncertainties in the stellar passband flux and radius ratio propagate into the inferred temperature of LP~261-75C. The ATMO 2020 temperature posterior is approximately Gaussian and centered at $1591$~K. Temperature is strongly correlated with the stellar passband flux, since a brighter stellar host requires greater brown-dwarf emission to reproduce the measured eclipse depth. The distribution on the radius ratio is approximately equal to its Gaussian prior set from the primary transit depth.

Panels (b) and (c) respectively display the ATMO 2020 chemical-equilibrium cooling tracks and the age posteriors for different evolutionary models at solar metallicity. In panel (b), the colored tracks represent fixed masses labeled in $M_{\mathrm{Jup}}$. Cooling is most rapid within the first gigayear, with the magnitude of the cooling slope, $|dT_{\mathrm{int}}/dt|$, decreasing at later ages. Thus, at fixed mass, the same temperature uncertainty leads to a larger age uncertainty for a cooler brown dwarf. The red region spans the $1\sigma$ mass and temperature constraints of LP 261-75C on the cooling tracks, corresponding to a median age of $1700$ Myr.

Panel (c) shows the age posterior distributions for the solar-metallicity models, with median ages spanning $1490$--$1940$~Myr. We visualize these distributions by applying a smoothed Gaussian kernel, with kernel standard deviations set to $15\%$ ($\approx 50-150 Myr$) of each posterior's standard deviation. All five distributions are asymmetric, with extended tails toward older ages, particularly for Diamondback and SM08. Approximately $17\%$ of the Diamondback samples and $25\%$ of the SM08 samples exceed $2500$~Myr. CBPD yields the youngest median age and notably the narrowest posterior, although the 68\% credible intervals of all five models overlap. This model age ordering differs from that of AD~3116B, for which the cloudy models yield the youngest ages.

\begin{figure}[ht!]
    \centering
    \includegraphics[width=\columnwidth]{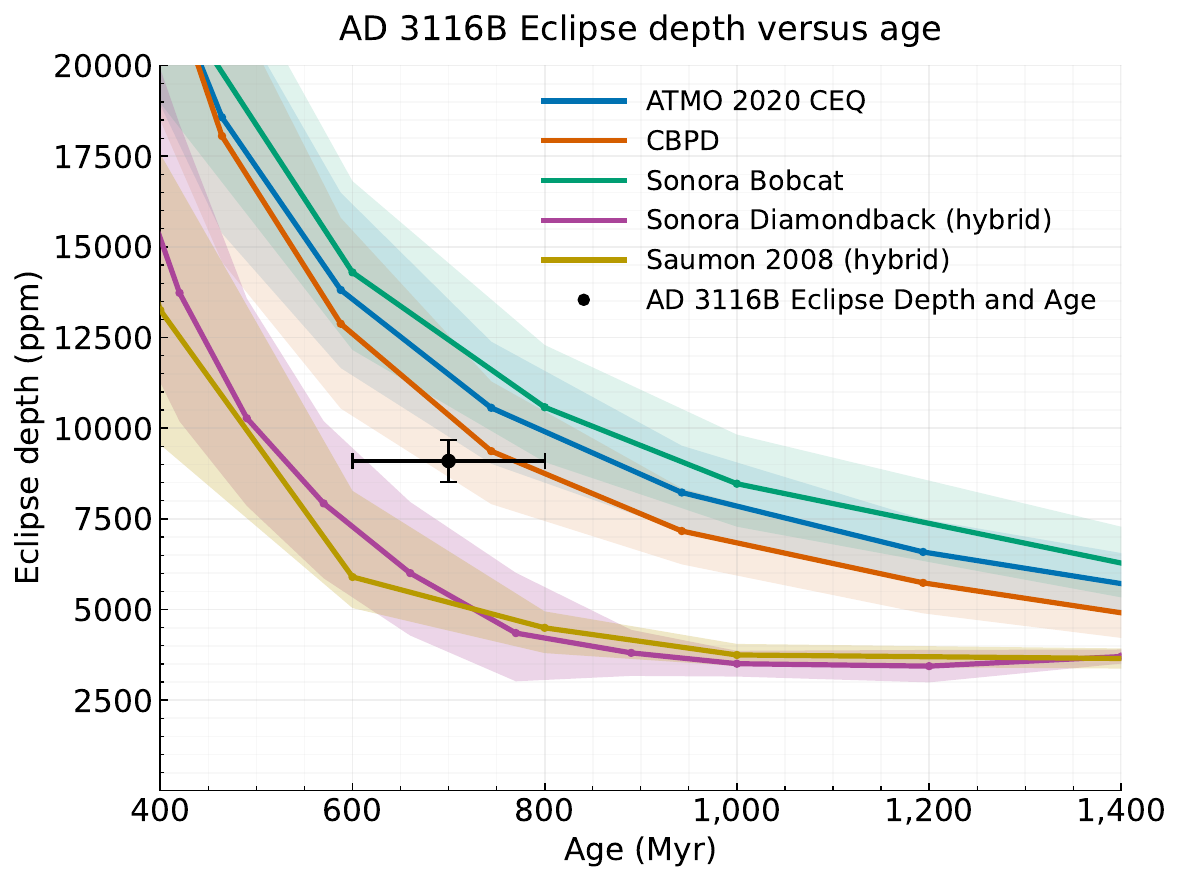}
    \caption{Predicted $J$-band eclipse depth of AD~3116B as a function of age for five evolutionary models. Shaded regions illustrate the uncertainty in predicted depth associated with the brown dwarf's mass uncertainty of $53.6 \pm 3.1\,M_{\mathrm{Jup}}$. The black point marks the inverse-variance-weighted eclipse depth, $\delta = 9060 \pm 560\,\mathrm{ppm}$. The horizontal error bar spans the adopted age range of 600--800\,Myr.}
    \label{fig:eclipse_depth_age}
\end{figure}

Figure~\ref{fig:eclipse_depth_age} illustrates this AD~3116B contrast through the predicted $J$-band eclipse depths for each evolutionary model. We calculate these depths using Equation~\ref{equation_eclipse_depth}, holding the stellar radius and flux fixed, because M dwarfs evolve slowly, while allowing the brown dwarf's radius and temperature to evolve. The resulting $(T_{\mathrm{int}}, \log g)$ values determine the spectra and thus the predicted $J$-band flux. Shaded regions show the uncertainties obtained by evaluating each cooling model at the $\pm1\sigma$ mass limits. We use the solar-metallicity SM08 and Diamondback hybrid models. For these two models, we interpolate the spectral flux between a cloudy boundary condition at 1400 K and a cloudless boundary condition at 1200 K, reflecting the L/T transition between a cloudy and cloudless atmosphere. At the central mass of $\approx53.6\,M_{\mathrm{J}}$, this temperature interval corresponds to ages of 1.0--3.0 Gyr for both cloudy models.

Figure~\ref{fig:eclipse_depth_age} also shows how the cooling solutions are bimodal. The three cloudless models follow broadly similar cooling trajectories, whereas the cloudy hybrid models predict younger ages for the observed eclipse depth and produce eclipses approximately half as deep near 900 Myr. Near 700 Myr, the CBPD prediction at the lower $1\sigma$ mass limit lies closest to the measured eclipse depth, while ATMO 2020 and Sonora Bobcat predict deeper eclipses. At the younger end of the adopted age interval, near 600 Myr, the Diamondback prediction at the upper mass limit overlaps the measurement. Diamondback therefore provides closer agreement near 600 Myr, whereas CBPD is favored toward 800 Myr in the age uncertainty range. All models exhibit substantial variation in eclipse depth across the mass uncertainty of $\sigma_{\rm M} = 3.1 \, M_{\rm Jup}$, highlighting the importance of precise mass measurements for distinguishing model predictions. The slight increase in the Diamondback eclipse depth toward 1400 Myr reflects the changing $J$-band flux as the spectral contribution shifts from cloudy to cloud-free conditions. A cloud-free spectrum at $T_{\mathrm{int}}=1200 \, \rm K$ can provide a $J$-band flux comparable to a cloudy $1400 \, \rm K$ spectrum, despite the cooler brown dwarf.

\subsection{Radius Evolutionary Constraints}

\begin{figure}[ht!]
    \centering
    \includegraphics[width=\columnwidth]{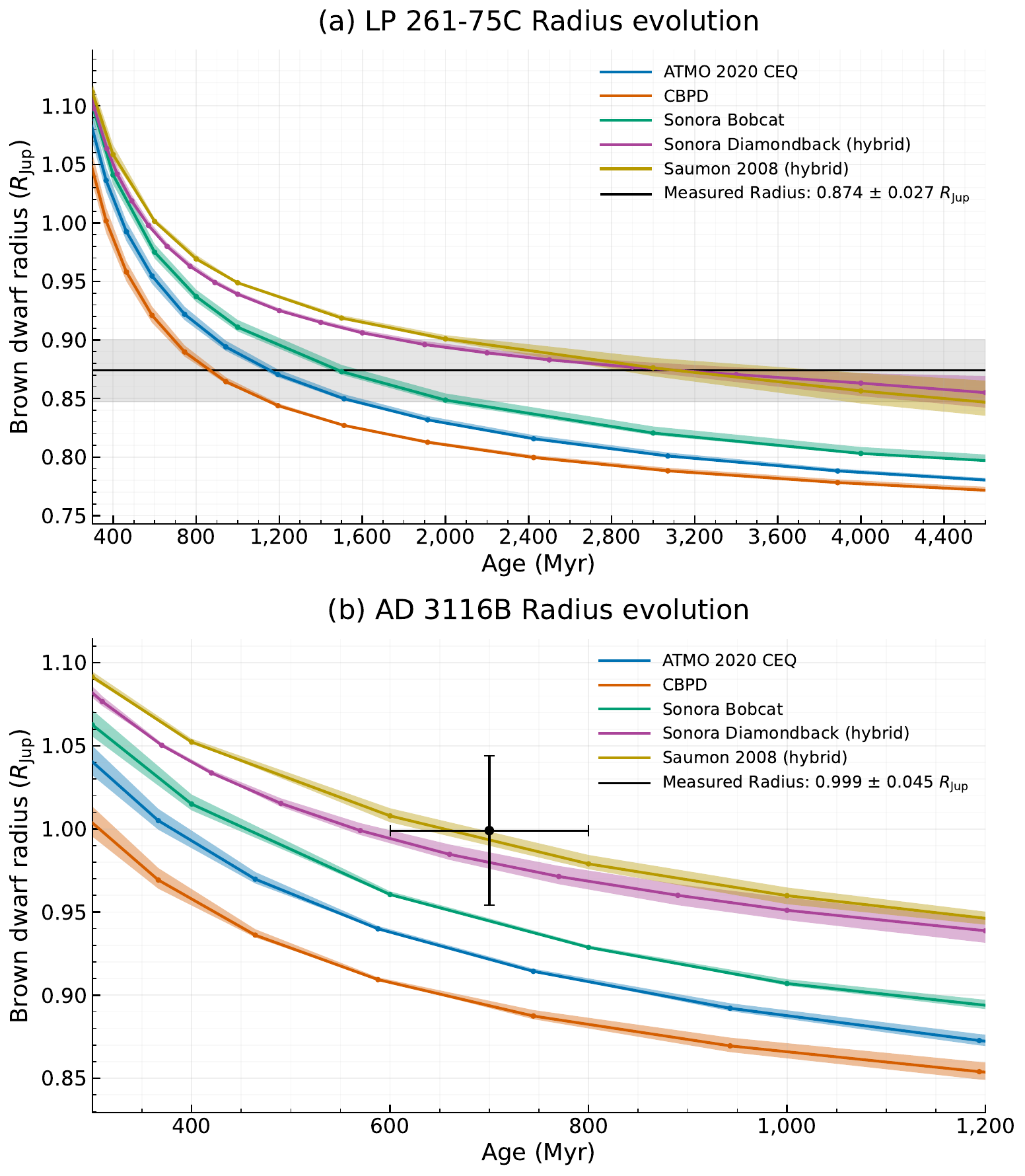}
    \caption{Radius evolution of (a) LP~261-75C and (b) AD~3116B for five evolutionary models. Solid curves correspond to the median measured masses, with coloured shading showing the mass uncertainties. The black line and grey band in panel~(a) indicate the measured radius and its uncertainty; the black point in panel~(b) includes radius and age uncertainties. Sonora Bobcat and Sonora Diamondback use solar metallicity, while Sonora Diamondback and SM08 use hybrid cooling.}    \label{fig:radius_evolution}
\end{figure}

As an additional evolutionary constraint, we compare the predicted brown dwarf radius as a function of age with the measured radius, allowing us to assess the consistency between ages inferred from radial contraction and the eclipse-based measurements. Figure~\ref{fig:radius_evolution} shows this comparison for LP~261-75C and AD~3116B. For LP~261-75C in panel (a), the median mass tracks intersect the measured radius, $R=0.874\pm0.027\,R_{\mathrm{Jup}}$, at ages of $860^{+280}_{-180}$, $1150^{+400}_{-260}$, and $1480^{+540}_{-350}\,\mathrm{Myr}$ for CBPD, ATMO~2020, and Sonora Bobcat, respectively. The Sonora Diamondback and SM08 hybrid models yield older ages of $3060^{+2120}_{-1320}$ and $3070^{+1440}_{-1090}\,\mathrm{Myr}$, with substantially larger uncertainties. This is because, near the measured radius, the cloudy models predict slower contraction (smaller \(|dR/dt|\)), translating the same radius uncertainty into a broader range of ages.

We can also compare the eclipse-based ages in Table~\ref{tab:evolutionary_comparison} with the radius-inferred ages. For LP 261-75C, CBPD, ATMO 2020, and Sonora Bobcat yield younger radius-inferred ages than the thermal cooling estimates, but only CBPD shows non-overlapping 68\% credible intervals. Conversely, the Diamondback hybrid and SM08 models yield older radius-inferred ages, although their considerably broad intervals overlap the eclipse-based estimates. As a consequence, the eclipse-based analysis provides a substantially tighter age constraint for the cloudy evolutionary models. For LP~261-75C, the eclipse-based ages and radius-evolution constraints both favour ages of order $1\,\mathrm{Gyr}$ or older, providing complementary evidence against membership in the much younger AB~Doradus moving group.

Panel~(b) of Figure~\ref{fig:radius_evolution} compares the measured radius of AD~3116B with evolutionary predictions over the adopted age range of $600$--$800\,\mathrm{Myr}$. The Sonora Diamondback and SM08 hybrid models reproduce the measured radius within $1\sigma$. At $700\,\mathrm{Myr}$, Sonora Bobcat, ATMO~2020, and CBPD instead underpredict the radius by approximately $1.2\sigma$, $1.7\sigma$, and $2.3\sigma$. These radius discrepancies are relevant to interpreting Figure~\ref{fig:eclipse_depth_age} because the eclipse-depth predictions use the model-predicted radii. Although CBPD provides the closest eclipse-depth prediction at the age of $700\,\mathrm{Myr}$, it predicts $R\simeq0.894\,R_{\mathrm{Jup}}$, substantially below the measured $R=0.999\pm0.045\,R_{\mathrm{Jup}}$, and produces the largest radius offset among the considered models. Diamondback provides a plausible simultaneous match to both observables near the younger end of the adopted age range, around $600\,\mathrm{Myr}$, for masses toward the upper end of the measured uncertainty range. Diamondback’s ability to account for both observables at the same age and mass makes it a particularly promising consideration for AD 3116B. Similarly, toward the older end of the adopted age range at $800\,\mathrm{Myr}$, ATMO 2020 and Sonora Bobcat can reproduce the observed eclipse depth to within $1\sigma$, while underpredicting the measured radius to within $2\sigma$.

\subsection{Mass of Wide-orbit Companion LP~261-75B}

Our eclipse-based system ages also allow us to estimate the mass of the wide companion LP~261-75B. As discussed in Section~\ref{sec:2pt1}, this directly imaged L6 brown dwarf lies at a projected separation of $450\pm120\,\mathrm{au}$. We combine its adopted effective temperature, $T_{\mathrm{eff}}=1500\pm150\,\mathrm{K}$, with the system-age posteriors inferred from LP~261-75C for $T_{\mathrm{eq}}=0$ to infer the companion mass. The quoted uncertainties below span the 16th--84th percentiles of the propagated mass distributions.

ATMO~2020 yields a mass of $62.4^{+6.2}_{-8.0}\,M_{\mathrm{Jup}}$, with CBPD and solar-metallicity Sonora Bobcat giving nearly identical estimates. Increasing the Bobcat metallicity to $[\mathrm{M/H}]=+0.5$ raises its inferred median mass by $3\,M_{\mathrm{Jup}}$, to $65.4^{+4.1}_{-6.2}\,M_{\mathrm{Jup}}$. The solar-metallicity Sonora Diamondback hybrid model instead yields $67.7^{+4.2}_{-6.1}\,M_{\mathrm{Jup}}$. At $[\mathrm{M/H}]=+0.5$, Diamondback gives a slightly narrower distribution and a median mass $1\,M_{\mathrm{Jup}}$ lower. All of these mass estimates have overlapping uncertainty intervals, despite the large differences between the different model-dependent ages. These mass estimates are substantially higher than the original estimate of $21^{+10}_{-5}\,M_{\mathrm{Jup}}$ obtained by assuming an age of $100$--$200\,\mathrm{Myr}$ \citep{reid2006lp26175}. These mass estimates are also similar to the measured mass of LP~261-75C, as the adopted $T_{\mathrm{eff}}=1500\pm150\,\mathrm{K}$ for LP~261-75B is quite similar to the eclipse-based temperatures of LP~261-75C.

An independent dynamical mass measurement would provide a valuable test of this evolutionary mass, but obtaining such a measurement for LP~261-75B is challenging. Adopting the projected separation of $450\,\mathrm{au}$ as an estimate of the semimajor axis, the total system mass implies an orbital period of $\sim15{,}000\,\mathrm{yr}$. Such a long period makes it difficult to constrain the orbital dynamics over observing baselines. For comparison, even long-period systems with dynamical mass constraints from radial velocities and astrometry, such as HD~19467B and HD~47127B, have orbital periods of 320 and 600 years, respectively \citep{hoch2024,bowler2021}.

\section{Future Prospects}\label{sec:future}

A natural question for future applications is whether our method can also constrain the ages of brown dwarfs orbiting Sunlike primaries. Compared with M-dwarf hosts, Sunlike primaries yield higher equilibrium temperatures at comparable orbital separations, so the observed brown dwarf flux can include a substantial contribution from atmospheric heating by absorbed starlight. Separating this contribution from the intrinsic emission requires constraints on the Bond albedo and the heat redistribution factor $f$, which can introduce strong degeneracies with $T_{\rm int}$ and consequently broaden the inferred age posterior. Additionally, the larger masses of Sunlike primaries can substantially enhance tidal heating at fixed orbital geometry (Equation~\ref{eq:tidal_temp}).

As an example, the CHEOPS study of KELT-1b, which orbits an F-type primary and has an equilibrium temperature of $\sim 2600$~K, combined observations in six optical and infrared passbands to infer $T_{\rm int}=2670^{+190}_{-150}$~K \citep{parviainen2022}. While the uncertainties of this measurement are small in an absolute sense, they translate to large uncertainties ($> 1$~Gyr) in age. Applying our technique to such systems would require irradiated brown dwarf atmosphere models \citep{mukherjee2023,hubeny2017} and evolutionary calculations that account for the slower interior cooling \citep{baraffe2003,mukherjee2026}. For instance, \citet{Beatty2018} investigated the emission of the brown dwarf CWW~89Ab, whose host system is a member of the $3.0\pm0.25$~Gyr-old open cluster Ruprecht~147, and found an overluminosity that is best explained by a dayside temperature inversion, which standard no-irradiation evolutionary models do not predict. Because of this sensitivity in how brown dwarf atmospheres respond to stellar heating, our method is best suited to cooler M-dwarf hosts, where these assumptions can be bounded.

Because transiting brown dwarfs are intrinsically rare, there are only a few systems for which we could currently expand our technique. Other known brown dwarfs transiting nearby M dwarfs include TOI-1278B \citep{artigau2021toi1278}, NLTT~41135B \citep{irwin2010nltt41135}, LHS~6343C \citep{johnson2011lhs6343}, TOI-2119B \citep{canas2022toi2119}, TOI-263B \citep{palle2021toi263}, and NGTS-7AB \citep{jackman2019ngts7}. Among these, TOI-1278B, LHS~6343C, and TOI-2119B stand out for their relatively long orbital periods ($> 7$~days), which minimize stellar irradiation even further and weaken tidal interactions that could have influenced the spin histories of their low-mass host stars. TOI-2119 is a particularly interesting system for calibrating spin down models given the 13.2-day rotation period of its M dwarf host, which \citet{newton2016mdwarf} showed to be uncommon relative to M dwarfs with rotation periods shorter than 10 days or longer than 70 days. Sadly, future missions are unlikely to expand this sample significantly. The ESA PLATO mission will search for transiting planets with an unprecedented precision \citep{rauer2025}, but with a bluer passband than TESS, it is unlikely to make significant gains for M dwarfs. The NASA Roman Space Telescope is expected to detect thousands of transiting companions around M dwarfs \citep{wilson2023}, but the host stars will be far too faint for ground-based follow-up observations.

Fortunately, our technique is not limited to transiting systems, and may be possible for dM+BD systems detected in Gaia DR4. \citet{lammers2026} predicted that the upcoming Gaia data release will contain the astrometric detections of thousands of giant planets with orbits between 1 and 3~au. The yield for brown dwarfs will likely be comparable; the astrometric signal of a companion scales in proportion to its mass, and brown dwarfs at these separations are roughly a tenth as common as their Jupiter-mass counterparts \citep[e.g.,][]{vanzandt2026}. If their masses are not already precisely constrained by the astrometry, brown dwarfs orbiting M dwarfs at these separations should be trivial targets for dynamical mass measurements due to their large radial velocity semi-amplitudes. The final observable needed to calculate stellar ages is the luminosity of the brown dwarf companion, which we will be able to measure in systems where the brown dwarf companions can be directly imaged. Upcoming ground-based instruments capable of high-resolution imaging in the near infrared, such as Keck/SCALES \citep{skemer2022}, may be able to achieve these measurements for relatively nearby systems. A dM+BD system 20~pc away with a separation of 2~au will have an angular separation of $0\farcs1$, and a relatively massive brown dwarf will have a modest contrast with its host that may be detectable with these upcoming instruments. \citet{kirkpatrick2024} reported $\sim 2,500$ M dwarfs within 20~pc of the Solar System; even if only $0.1\%$ of these M dwarfs have brown dwarf companions with detected astrometric orbits/accelerations, it would add several dozen new tidally detached systems with which we could calibrate spin down models for cool stars.

\section{Conclusions}\label{sec:conclusions}
We present a framework for estimating the ages of systems with M dwarfs and transiting brown dwarf companions. Using a suite of brown dwarf evolutionary models, we estimate ages for the LP~261-75 ($1.3-2.9$~Gyr) and AD~3116 ($0.4-1.3$~Gyr) systems. The latter of these two is a rare benchmark system with an independently measured age of $0.6-0.8$~Gyr, based on membership in the Praesepe cluster, allowing us to demonstrate the general accuracy of our technique. We perform an in-depth exploration of how the different physics included in the different models (e.g., equations of state, atmospheric metallicity, and clouds) impact our inferred ages. Lastly, we use our new age estimate to provide an updated mass estimate for the wide-separation brown dwarf companion LP~261-75B, finding that the companion is likely $3\times$ more massive than earlier estimates. Our technique provides a potential pathway for measuring the ages of more M dwarfs in the near future, specifically by targeting those with brown dwarf companions detected by Gaia. Such measurements would provide new constraints on M dwarf evolution, enabling new investigations of stellar interiors, galactic archaeology, and star-planet interactions.


\begin{acknowledgments}

A.G. was supported by Princeton University's Office of Undergraduate Research through the Undergraduate Fund for Academic Conferences and the Banes Family Senior Thesis Endowment Fund. S.G. was supported by an NSF Astronomy and Astrophysics Postdoctoral Fellowship under award AST-2303922. We thank the anonymous referee for their insightful comments and constructive feedback, which helped improve the manuscript. We thank ZJ Zhang for providing the IRTF/SpeX spectrum of LP~261-75A. Generative AI tools were used to assist with figure preparation and code debugging. They were not used for data analysis or writing the manuscript. 

\end{acknowledgments}





%
\facilities{Hale (WIRC), IRTF (SpeX), Keck:I (HIRES), Gemini:Gillett (MAROON-X), FLWO:1.5m (TRES), SPHEREx, TESS, Kepler}

\software{\texttt{dynesty} \citep{speagle2020}, \texttt{emcee} \citep{foreman2013}, \texttt{exowirc} \citep{vissapragada2020exowirc}, \texttt{PyMC3} \citep{salvatier2016probabilistic}, \texttt{juliet} \citep{espinoza2019}}

\newpage
\appendix

\section{Global Fit Residuals}

\begin{figure}[ht!]
    \centering
    \includegraphics[width=\textwidth]{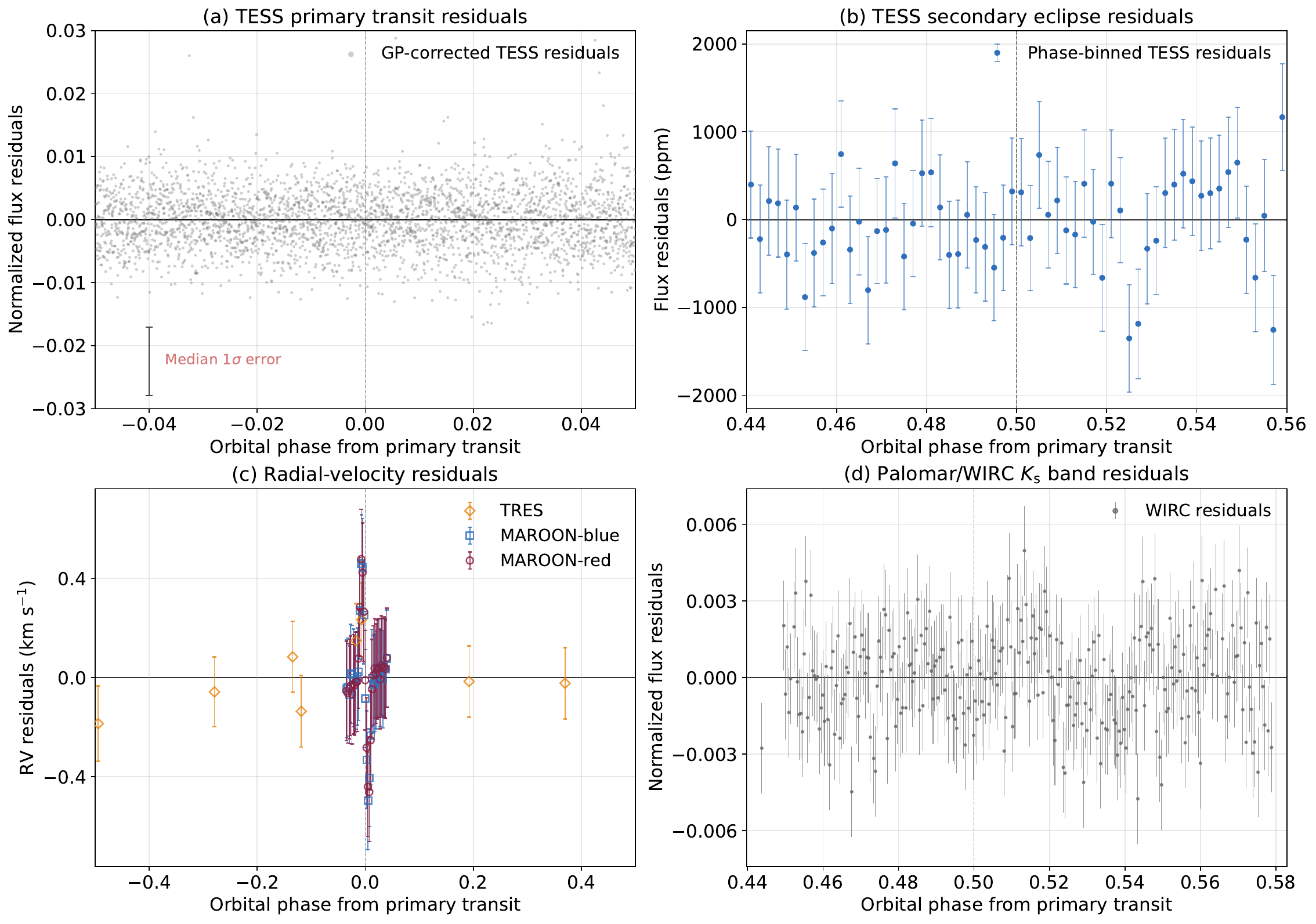}
    \caption{LP~261-75C global fit residuals corresponding to the data in Figure~\ref{fig:lp261_joint_fit}.}
    \label{fig:lp261_residuals}
\end{figure}

\begin{figure}[ht!]
    \centering
    \includegraphics[width=\textwidth]{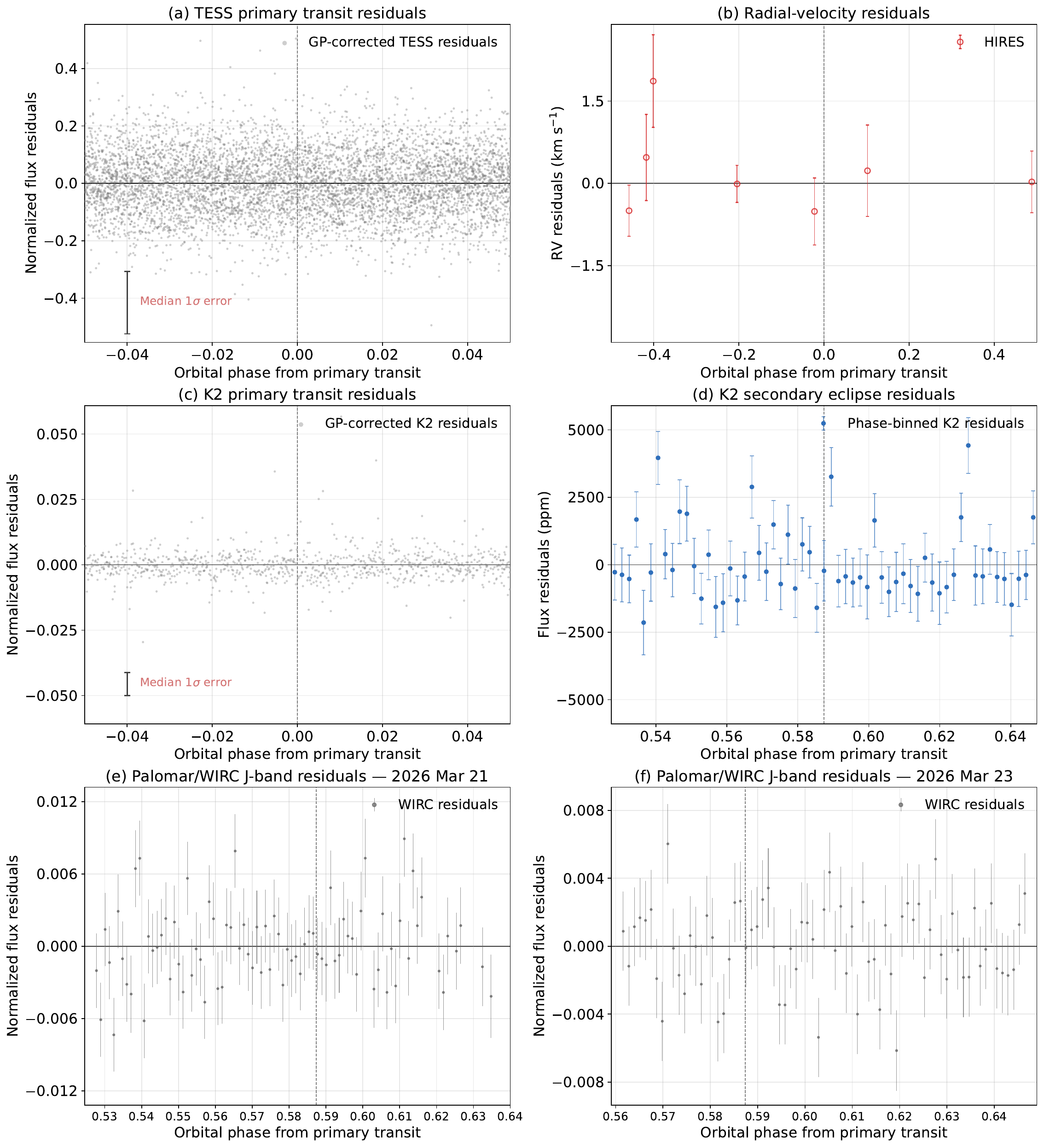}
    \caption{AD~3116B global fit residuals corresponding to the data in Figure~\ref{fig:ad3116b_joint_fit}.}
    \label{fig:ad3116_residuals}
\end{figure}

Figures~\ref{fig:lp261_residuals} and ~\ref{fig:ad3116_residuals} display the residuals to the global fits shown in Figures~\ref{fig:lp261_joint_fit} and \ref{fig:ad3116b_joint_fit}, respectively. We show the residuals here rather than earlier to reduce figure crowding.

\clearpage


\bibliography{references}{}
\bibliographystyle{aasjournalv7.1}



\end{document}